# The Lyapunov Concept of Stability from the Standpoint of Poincare's Approach: General Procedure of Utilization of Lyapunov Functions for Non-Linear Non-Autonomous Parametric Differential Inclusions

Myroslav Sparavalo
NYC Transit Authority, Telecommunications Division
126 53rd Street, NY, NY 10019, USA
mksparavalo@yahoo.com

***In memory of Henri Poincare and Aleksandr Lyapunov***

**ABSTRACT**

The objective of the research is to develop a general method of constructing Lyapunov functions for non-linear non-autonomous differential inclusions described by ordinary differential equations with parameters. The goal has been attained through the following ideas and tools. First, three-point Poincare's strategy of the investigation of differential equations and manifolds has been used. Second, the geometric-topological structure of the non-linear non-autonomous parametric differential inclusions has been presented and analyzed in the framework of hierarchical fiber bundles. Third, a special canonizing transformation of the differential inclusions that allows to present them in special canonical form, for which certain standard forms of Lyapunov functions exist, has been found. The conditions establishing the relation between the local asymptotical stability of two corresponding particular integral curves of a given differential inclusion in its initial and canonical forms are ascertained. The global asymptotical stability of the entire free dynamical systems as some restrictions of a given parametric differential inclusion and the whole latter one per se has been investigated in terms of the classificational stability of the typical fiber of the metabundle. There have discussed the prospects of development and modifications of the Lyapunov second method in the light of the discovery of the new features of Lyapunov functions.

## INTRODUCTION

Stability together with controllability on many occasions composes the core of research interests in the modern science and technologies where the mathematical models of physical phenomena are used to break new ground. Today even in medicine the process of successful healing and recovering is described by mathematical models of living organisms able despite the influence of harmful external perturbations to return to their initial healthy state. Sometimes to achieve the complete curing it is necessary to solve the inverse problem of stability namely, to make damaged genes lose the stability of the transmission, regeneration and restoration of their "nefarious properties". Unfortunately high dimensionalities, nonlinearity, parametric and structural uncertainty of mathematical models create a lot of troubles on the way to solving problems of stability. In fact by now there is only one method that allows to tackle the problems under such tough conditions in the most general case. This is Lyapunov second method or Lyapunov functions. But since the time, that is about a hundred years ago, when A. Lyapunov introduced the special functions later named after him in the theory of stability the problem of finding or constructing them in the most general case has seemed desperately insurmountable. There is no wonder because one does not exist any general method of finding the analytical solutions of the systems of nonlinear non-autonomous differential equations. This fact has endowed Lyapunov functions with a sort of mystery. It has been unclear how Lyapunov functions relate to the first integrals of the systems of differential equations if they do in general. They have always produced the impression of some kind of artificiality imposed on the systems. It has been considered that even if having mostly extraneous rather than the innate relation to the systems, Lyapunov functions may be interjected with or interpreted as part of their intrinsic essence. And mathematicians and engineers have had to put up with this uncertainty for almost a century. However it transpires that the Henri Poincare's strategy of the investigation of differential equations and manifolds allows us to handle successfully even such hopeless problems.

It would be expedient to remind his approach.

1. Henri Poincare preferred the non-parametric form of the manifold representation to parametric one namely, a *m*-dimensional submanifold is defined by a system of *l* implicit functional equations, each of which depends of *n* variables and describes the corresponding one-codimensional or $(n - 1)$-dimensional manifold. The following relation holds true in this case: $m = n - l$. Thus the *m*-dimensional submanifold can be represented by the intersection of *l* of the one-codimensional manifolds.
2. He suggested to carry out the investigation of *m*-dimensional submanifold beginning with the study of each of *l* of the one-codimensional manifolds, which, intersecting, form it. Why? Since as a part of the latter ones the former inherits their properties in certain way.
3. The right-hand sides of the differential equations completely define the qualitative and quantitative behavior the dynamical systems they describe.

Now let us formulate the objectives of the proposed research.

First, we have intentions to establish the relation between Lyapunov functions and the intrinsic properties of the systems of differential equations carried by their right-hand sides and/or first integrals. Second, we would like to develop the general method of constructing Lyapunov functions regardless of the dimensionality, all kinds of nonlinearity and autonomousness of the dynamic systems. In order to expand the range of the generality of the investigated systems of ordinary differential equations not containing controlling functions in their right-hand sides to the possible maximum we will consider parametric differential inclusions.

Our plan of actions is the following one.

First, we will give certain geometric-topological representation to the nonlinear dynamics of parametric differential inclusions, where the center of investigative interest will be integral curves and invariant manifolds of various dimensions. This goal will be attained in the framework of fiber bundles and foliations. Second, we will find a special transformation of the initial parametric differential inclusions that allows us to render it certain “convenient and universal” form to handle the problem of stability further. The “convenience and universality” should be understood in the following sense. Figuratively, the stability can be explained as the ability of systems to produce the limited and even tending to zero with time system reaction to the various input disturbances ranging from small to large magnitudes. It is the wide variety of input disturbances together with nonlinearity are the principal real plagues in the problem of finding and constructing Lyapunov functions. The special “convenient” transformation is meant to “tuck” all these plagues into the right-hand sides of the transformed parametric differential inclusions. Third, we will establish certain correspondence of stability between the initial and transformed parametric differential inclusions in the following sense: if the latter ones behave as asymptotically stable then the former ones also do. Then we will find the Lyapunov functions of so-called canonical form that are universal for all the class and types of the transformed parametric differential inclusions. Forth, the last thing we plan to do is to obtain the conditions for the local asymptotical stability of integral curves, the global asymptotical stability for dynamic systems and parametric differential inclusions, for which the dynamic systems are some restrictions when the vector of parameters of the inclusion assumes certain values.

## 1. BASIC GEOMETRIC-TOPOLOGICAL STRUCTURES OF PARAMETRIC DIFFERENTIAL INCLUSIONS IN THE FRAMEWORK OF FIBER BUNDLES

Consider a differential inclusion as follows

$$\frac{dx}{dt} \in \bigcup_{\xi \in \Xi} f\left(t, x; \xi\right), \tag{1}$$

where $t \in \left[t_0; +\infty\right] = T \subset R_t^1$ is time and for simplicity we will consider the ray $T$ unchangeable; $x = \left(x_1, ... x_n\right) \in R_x^n$ is a phase vector; $R_t^1$ and $R_x^n$ are Euclidean phase spaces with Cartesian rectangular coordinate systems; $\xi = \left(\xi_1, ..., \xi_m\right) \in \Xi \subseteq R_\xi^m$ is a vector of parameters, which values belong to some compact open parameter manifold $\Xi$ with $\dim \Xi = m$, $m \geq n$; $f\left(t, x; \xi\right) =$ $= \left(f_1\left(t, x; \xi\right), ..., f_n\left(t, x; \xi\right)\right) \in C^r$ is a vector-function of indicatrix field; $\left(t, x; \xi\right) \in T \times R_x^n \times \Xi$; $r \geq 1$.

Let us make some introductory conventions and designations.

Denote $\xi_0 = \left(\xi_{1,0}, ..., \xi_{m,0}\right) \in \Xi$ some given nominal or distinguished point of the vector of parameters $\xi$. The necessity of the introduction of the point can be substantiated by considering its neighborhood $\Xi_\xi \subseteq \Xi$, when we tackle problems of local stability. Make the convention that the mathematical symbol “hat” or “^” denotes some specific value of any variable or vector considered in the paper. For example, $\hat{\xi}_0$ designates some specific value of the distinguished point $\xi_0 \in \Xi$, $\hat{\xi} \neq \hat{\xi}_0$ is any concrete value of the vector $\xi$ different from the distinguished $\hat{\xi}_0$;

$\hat{x}_0 = \left(\hat{x}_{1,0},...\hat{x}_{n,0}\right)$ denotes a specific value of the initial point $x_0$ of the phase vector $x$ and $\hat{x} = \left(\hat{x}_1,...\hat{x}_n\right)$ is some concrete value of $x$ different from $\hat{x}_0$. Introduce the manifold $X_{t_0} \subseteq R_x^n$ containing all the initial points $x_0 = \left(x_{1,0},...x_{n,0}\right)$ of the phase vector $x$. We will call $X_{t_0}$ the manifold of the initial points of the phase vector $x$.

If we fix the value of the vector of parameter $\xi = \hat{\xi}$ then the differential inclusion (1) turns into the following free dynamic system

$$\frac{dx}{dt} = f\left(t, x; \hat{\xi}\right), \tag{2}$$

where the system (2) can be considered the restriction of the differential inclusion (1) to $T \times R_x^n$.

Let

1. $x = x\left(t; \hat{x}_0, \hat{\xi}\right) = \left(x_1\left(t; \hat{x}_0, \hat{\xi}\right),..., x_n\left(t; \hat{x}_0, \hat{\xi}\right)\right)$ be a particular solution to the system of ordinary differential equations (2) over $t \in T$ and at $x_0 = \hat{x}_0 \in X_{t_0}$, which also denotes the specific phase trajectory or orbit as of the dynamical system (2) so of the differential inclusion (1) with the concrete initial point of the phase vector $\hat{x}_0 = \left(\hat{x}_{1,0},...\hat{x}_{n,0}\right) \in X_{t_0}$ and the concrete value $\hat{\xi} = \left(\hat{\xi}_1,...,\hat{\xi}_m\right) \in \Xi$ of the vector of parameters $\xi$.
2. $x = x\left(t; x_0, \hat{\xi}\right) = \left(x_1\left(t; x_0, \hat{\xi}\right),..., x_n\left(t; x_0, \hat{\xi}\right)\right)$ be a general solution to the dynamical system (2) with $\forall x_0 \in X_{t_0}$ and $\hat{\xi} \in \Xi$.
3. $x = x\left(t; x_0, \xi\right) = \left(x_1\left(t; x_0, \xi\right),..., x_n\left(t; x_0, \xi\right)\right)$ be a general solution to the differential inclusion (1) with $\forall x_0 \in X_{t_0}$ and $\xi \in \Xi$.

Since we deal with non-autonomous systems it would be expedient to introduce the integral curve $x_t\left(\hat{x}_0, \hat{\xi}\right) = \left(t, x\left(t; \hat{x}_0, \hat{\xi}\right)\right)$ corresponding to the phase trajectory $x\left(t; \hat{x}_0, \hat{\xi}\right)$. We define the extended phase space or motion space as given below

$$X_t\left(\hat{\xi}\right) = \bigcup_{\forall \hat{x}_0 \in X_{t_0}} x_t\left(\hat{x}_0, \hat{\xi}\right). \tag{3}$$

In the designation $X_t\left(\hat{\xi}\right)$ the symbol $\hat{\xi}$ says that the motion space $X_t\left(\hat{\xi}\right)$ of the dynamic system (2) is just a section of some more total space

$$X_t(\xi)=\bigcup_{\forall\hat{\xi}\in\Xi}X_t\left(\hat{\xi}\right) \tag{4}$$

of the differential inclusion (1) at $\xi=\hat{\xi}$.

Further in our reasoning we will use the terminology and definitions of topological and geometrical objects and constructions according to [1].

In fact, the $x_t\left(\hat{x}_0,\hat{\xi}\right)$ represents a typical fiber of the fiber bundle of integral curves

$$\left\{\Gamma_x\left(t,x;x_0,\xi\right),\pi_x\left(x_0,\xi\right),X_{t_0}\times\Xi,\mathrm{G}_{\hat{x}_0,\hat{\xi}}\left(t,x\right)\right\}, \tag{5}$$

where $\Gamma_x\left(t,x;x_0,\xi\right)=\left\{x_t\left(x_0,\xi\right)=\left(t,x\left(t;x_0,\xi\right)\right),\left(t,x\right)\in T\times R_x^n;\forall\left(x_0,\xi\right)\in X_{t_0}\times\Xi\right\}$ is the total space being a $\left(n+1;n,m\right)$-dimensional smooth manifold with the projection $\pi_x\left(x_0,\xi\right)$: $\Gamma_x\left(t,x;x_0,\xi\right)\to X_{t_0}\times\Xi$ ***(we call it the total projection)***, the base space $X_{t_0}\times\Xi$ ***(we call it the total base space)***, the typical fiber $\Gamma_{\hat{x}_0}\left(t,x;\hat{x}_0,\hat{\xi}\right)=\left\{x_t\left(\hat{x}_0,\hat{\xi}\right)=\left(t,x\left(t;\hat{x}_0,\hat{\xi}\right)\right)\in X_t\left(\hat{\xi}\right)\right\}\cong$ $\cong\pi_x^{-1}\left(\hat{x}_0,\hat{\xi}\right)$ ***(we call it the total typical fiber)*** and the one-parameter trivial structure (Lie) group $\mathrm{G}_{\hat{x}_0,\hat{\xi}}\left(t,x\right)$ ***(we call it the total structure group)*** acting on the typical fiber $\Gamma_{\hat{x}_0}\left(t,x;\hat{x}_0,\hat{\xi}\right)$.

**Definition 1.** A fiber bundle is called the fiber metabundle if its typical fiber or/and base space is the total space of another fiber bundle, which is called the fiber subbundle.

Now let us show that the fiber bundle (5) $\left\{\Gamma_x\left(t,x;x_0,\xi\right),\pi_x\left(x_0,\xi\right),X_{t_0}\times\Xi,\mathrm{G}_{\hat{x}_0,\hat{\xi}}\left(t,x\right)\right\}$ can be represented as a fiber metabundle.

Really, it is obvious that the total space ***(we call it the total subspace 1)*** $\Gamma_x\left(t,x;\hat{x}_0,\xi\right)=$

$$=\left\{x_t\left(\hat{x}_0,\xi\right)=\left(t,x\left(t;\hat{x}_0,\xi\right)\right)\forall\xi\in\Xi\Leftrightarrow\bigcup_{\forall\hat{\xi}\in\Xi}x_t\left(\hat{x}_0,\hat{\xi}\right),\left(t,x\right)\in T\times R_x^n;\hat{x}_0\in X_{t_0}\right\}$$

with the projection $\pi_x\left(\hat{x}_0,\xi\right)$ ***(we call it the section projection 2)***, the base space $\Xi$ ***(we call it the base subspace)*** and the one-parameter trivial structure (Lie) group $\mathrm{G}_{\hat{x}_0,\hat{\xi}}\left(t,x\right)$ acting on the typical fiber $\Gamma_{\hat{x}_0}\left(t,x;\hat{x}_0,\hat{\xi}\right)=\left\{x_t\left(\hat{x}_0,\hat{\xi}\right)\in X_t\left(\hat{\xi}\right)\right\}\cong\pi_x^{-1}\left(\hat{x}_0,\hat{\xi}\right)$ can be considered a fiber bundle or fiber subbundle in the structure of the fiber metabundle (5) as follows

$$\left\{\Gamma_x\left(t,x;\hat{x}_0,\xi\right),\pi_x\left(\hat{x}_0,\xi\right),\Xi,\mathrm{G}_{\hat{x}_0,\hat{\xi}}\left(t,x\right)\right\}. \tag{6}$$

Analogously,

$$\left\{\Gamma_x\left(t,x;x_0,\hat{\xi}\right),\pi_x\left(x_0,\hat{\xi}\right),X_{t_0},\mathrm{G}_{\hat{x}_0,\hat{\xi}}\left(t,x\right)\right\} \tag{7}$$

can be considered a fiber bundle or fiber subbundle in the structure of the fiber metabundle (5) with $\Gamma_x\left(t,x;x_0,\hat{\xi}\right)=\left\{X_t\left(\hat{\xi}\right)=\bigcup_{\forall\hat{x}_0\in X_{t_0}}x_t\left(\hat{x}_0,\hat{\xi}\right),\left(t,x\right)\in T\times R_x^n;\hat{\xi}\in\Xi\right\}$ as the total space ***(we call it the total subspace 2)***, $\pi_x\left(x_0,\hat{\xi}\right)$ as the projection ***(we call it the section projection 1)***, $X_{t_0}$ as the base space ***(we call it the base subspace)*** and the one-parameter structure group $\mathrm{G}_{\hat{x}_0,\hat{\xi}}\left(t,x\right)$, which is trivial and acts on the typical fiber $\Gamma_{\hat{x}_0}\left(t,x;\hat{x}_0,\hat{\xi}\right)=\left\{x_t\left(\hat{x}_0,\hat{\xi}\right)\in X_t\left(\hat{\xi}\right)\right\}\cong\pi_x^{-1}\left(\hat{x}_0,\hat{\xi}\right)$ being equal to the total typical fiber of the (5) and the typical fiber of (6).

Now we are able to describe the fiber bundle (5) $\left\{\Gamma_x\left(t,x;x_0,\xi\right),\pi\left(x_0,\xi\right),X_{t_0}\times\Xi,\mathrm{G}_{x_0,\xi}\left(t,x\right)\right\}$ in two ways as

1) a fiber metabundle

$$\left\{\Gamma_x\left(t,x;x_0,\xi\right),\pi_x\left(x_0,\hat{\xi}\right),X_{t_0},\mathrm{G}_{\hat{x}_0,\xi}\left(t,x\right)\right\} \tag{8}$$

with the total metabundle space $\Gamma_x\left(t,x;x_0,\xi\right)=\left\{x_t\left(x_0,\xi\right)=\left(t,x\left(t;x_0,\xi\right)\right),\left(t,x\right)\in T\times R_x^n;\right.$ $\left.\forall\left(x_0,\xi\right)\in X_{t_0}\times\Xi\right\}$, the section projections 1 $\pi_x\left(x_0,\hat{\xi}\right)$, the total base space $X_{t_0}$, and the $m$ -parameter structure group $\mathrm{G}_{\hat{x}_0,\xi}\left(t,x\right)$, which acts on the typical fiber $\Gamma_x\left(t,x;\hat{x}_0,\xi\right)=$ $=\left\{X_t^{\xi}\left(\hat{x}_0\right)=\bigcup_{\forall\hat{\xi}\in\Xi}x_t\left(\hat{x}_0,\hat{\xi}\right),\left(t,x\right)\in T\times R_x^n;\hat{x}_0\in X_{t_0}\right\}\cong\pi_x^{-1}\left(\hat{x}_0,\xi\right)$ being the total subspace 1 of the fiber bundle (6), where $X_t^{\xi}\left(\hat{x}_0\right)=\bigcup_{\forall\hat{\xi}\in\Xi} x_t\left(\hat{x}_0,\hat{\xi}\right)$ is a bouquet of the integral curves starting from the initial point $\hat{x}_{t_0}=\left(t_0,\hat{x}_0\right)$.

2) a fiber metabundle

$$\left\{\Gamma_x\left(t,x;x_0,\xi\right),\pi_x\left(\hat{x}_0,\xi\right),\Xi,\mathrm{G}_{x_0,\hat{\xi}}\left(t,x\right)\right\} \tag{9}$$

with the total metabundle space $\Gamma_x\left(t,x;x_0,\xi\right)=\left\{x_t\left(x_0,\xi\right)=\left(t,x\left(t;x_0,\xi\right)\right),\left(t,x\right)\in T\times R_x^n;\ \forall\left(x_0,\xi\right)\in X_{t_0}\times\Xi\right\}$,the section projection 2 $\pi_x\left(\hat{x}_0,\xi\right)$, the total base space $\Xi$ and the $n$-parameter structure group $\mathrm{G}_{x_0,\hat{\xi}}\left(t,x\right)$, which acts on the typical fiber $\Gamma_x\left(t,x;x_0,\hat{\xi}\right)=$

$=\left\{X_t\left(\hat{\xi}\right)=\bigcup_{\forall\hat{x}_0\in X_{t_0}} x_t\left(\hat{x}_0,\hat{\xi}\right),\left(t,x\right)\in T\times R_x^n;\hat{\xi}\in\Xi\right\}\cong\pi_x^{-1}\left(x_0,\hat{\xi}\right)$being the total subspace 2 of the fiber bundle (7). Fig. 1 illustrates the structure of this fiber metabundle.

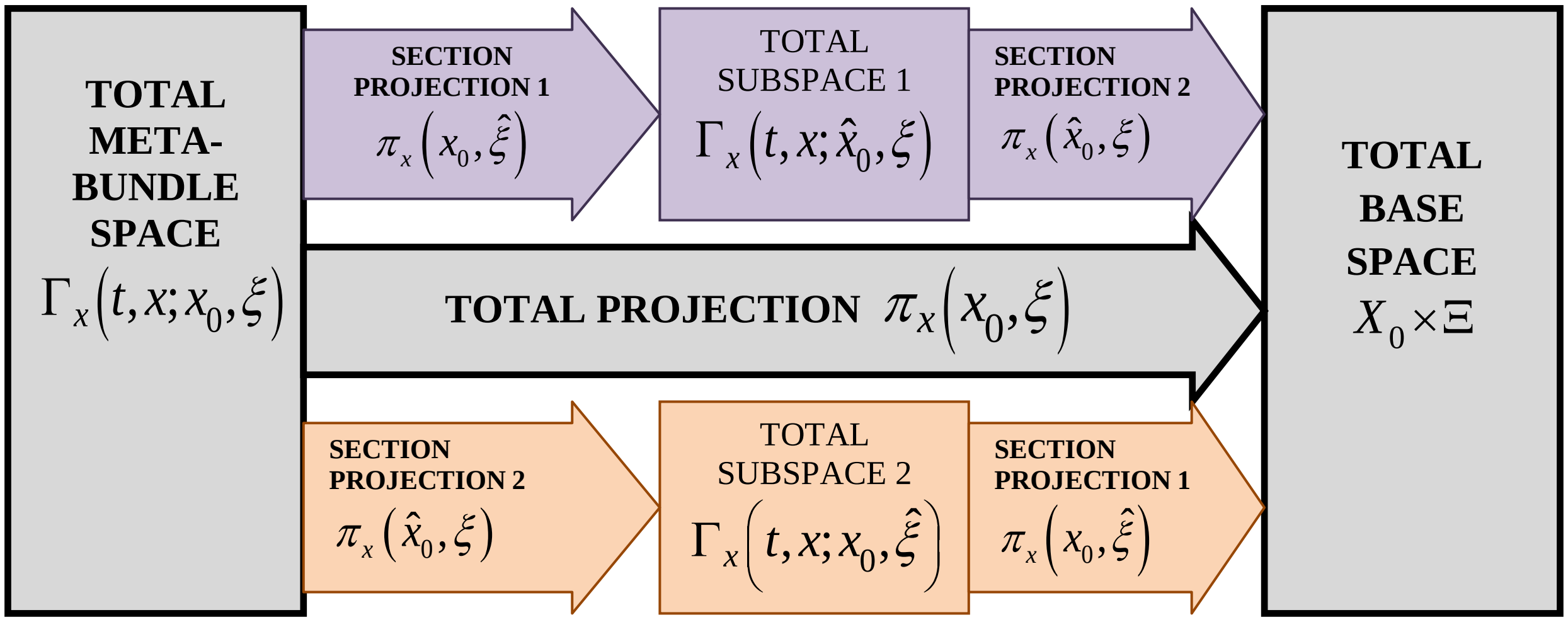

**FIG. 1**

Thus the fiber metabundle (5) $\left\{\Gamma_x\left(t,x;x_0,\xi\right),X_0\times\Xi,\pi\left(x_0,\xi\right),\mathrm{G}_{\hat{x}_0,\hat{\xi}}\left(t,x\right)\right\}$ decomposes hierarchically in two ways, namely

i. The inverse section projection 1 $\pi_x\left(x_0,\hat{\xi}\right)$ binds up all its typical fibers of the form $\Gamma_x\left(t,x;\hat{x}_0,\xi\right)=\left\{X_t^{\xi}\left(\hat{x}_0\right)=\bigcup_{\forall\hat{\xi}\in\Xi}x_t\left(\hat{x}_0,\hat{\xi}\right),\left(t,x\right)\in T\times R_x^n;\hat{x}_0\in X_{t_0}\right\}\cong\pi_x^{-1}\left(\hat{x}_0,\xi\right)$ being the bouquets of the integral curves starting from the same initial points, where $x_0$ runs over all the points of the base subspace $X_{t_0}$, in the total metabundle space $\Gamma_x\left(t,x;x_0,\xi\right)$ of the metabouquet or fiber metabundle (5) $\left\{\Gamma_x\left(t,x;x_0,\xi\right),\pi_x\left(x_0,\xi\right),X_{t_0}\times\Xi,\mathrm{G}_{\hat{x}_0,\hat{\xi}}\left(t,x\right)\right\}$. But in its turn, $\Gamma_x\left(t,x;\hat{x}_0,\xi\right)$ is the total subspace of the intermediate-level fiber subbundle (6) $\left\{\Gamma_x\left(t,x;\hat{x}_0,\xi\right),\pi_x\left(\hat{x}_0,\xi\right),\Xi,\mathrm{G}_{\hat{x}_0,\hat{\xi}}\left(t,x\right)\right\}$ with the base subspace $\Xi$, the section projection 2 $\pi_x\left(\hat{x}_0,\xi\right)$, the structural group $\mathrm{G}_{\hat{x}_0,\hat{\xi}}\left(t,x\right)$ acting on the typical fiber $\Gamma_{\hat{x}_0}\left(t,x;\hat{x}_0,\hat{\xi}\right)=$ $=\left\{x_t\left(\hat{x}_0,\hat{\xi}\right)\in X_t\left(\hat{\xi}\right)\right\}\cong\pi_x^{-1}\left(\hat{x}_0,\hat{\xi}\right)$.

ii. The inverse section projection 2 $\pi_x\left(\hat{x}_0,\xi\right)$ binds up all its typical fibers $\Gamma_x\left(t,x;x_0,\hat{\xi}\right)=\left\{X_t\left(\hat{\xi}\right)=\bigcup_{\forall\hat{x}_0\in X_{t_0}}x_t\left(\hat{x}_0,\hat{\xi}\right),\left(t,x\right)\in T\times R_x^n;\hat{\xi}\in\Xi\right\}\cong\pi_x^{-1}\left(x_0,\hat{\xi}\right)$being the sheaves of integral curves in the form of motion spaces, where $\xi$ runs over all the points $\hat{\xi}$ of the base subspace $\Xi$, into the total metabundle space $\Gamma_x\left(t,x;x_0,\xi\right)$of the metasheaf or fiber metabundle (5)$\left\{\Gamma_x\left(t,x;x_0,\xi\right),\pi_x\left(x_0,\xi\right),X_{t_0}\times\Xi,\mathrm{G}_{\hat{x}_0,\hat{\xi}}\left(t,x\right)\right\}$. But in its turn $\Gamma_x\left(t,x;x_0,\hat{\xi}\right)$ is the total subspace of the intermediate-level fiber subbundle (7) $\left\{\Gamma_x\left(t,x;x_0,\hat{\xi}\right),\pi_x\left(x_0,\hat{\xi}\right),\mathrm{X}_{t_0},\mathrm{G}_{\hat{x}_0,\hat{\xi}}\left(t,x\right)\right\}$ with the base subspace $X_{t_0}$, the section projection 1 $\pi_x\left(x_0,\hat{\xi}\right)$, the structural group $\mathrm{G}_{\hat{x}_0,\hat{\xi}}\left(t,x\right)$ acting on the typical fiber $\Gamma_{\hat{x}_0}\left(t,x;\hat{x}_0,\hat{\xi}\right)=\left\{x_t\left(\hat{x}_0,\hat{\xi}\right)\in X_t\left(\hat{\xi}\right)\right\}\cong\pi_x^{-1}\left(\hat{x}_0,\hat{\xi}\right)$.

There is one more important thing we need to add to our reasoning is that the composition of section projections 1 and 2 is commutative and isomorphic on the fiber metabundle, namely

$$\pi_x\left(x_0,\hat{\xi}\right)\circ\pi_x\left(\hat{x}_0,\xi\right)\cong\pi_x\left(\hat{x}_0,\xi\right)\circ\pi_x\left(x_0,\hat{\xi}\right)\cong\pi_x\left(x_0,\xi\right) \tag{10}$$

The hierarchical structure of fiber metabundle generated by the inclusion (1) is shown on Fig. 2.

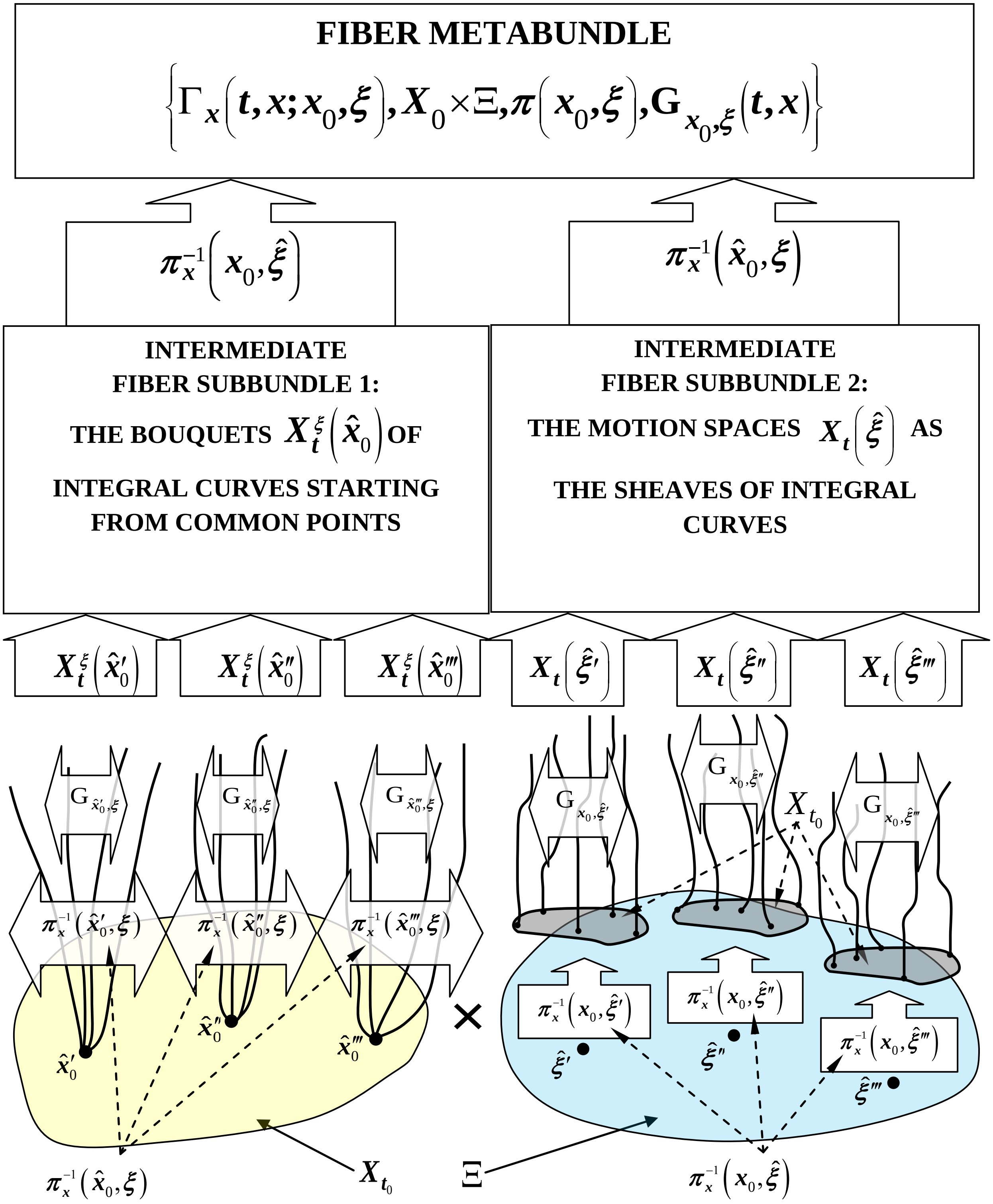

FIBER METABUNDLE
$\left\{\Gamma_x\left(t,x;x_0,\xi\right),X_0\times\Xi,\pi\left(x_0,\xi\right),G_{x_0,\xi}\left(t,x\right)\right\}$
$\pi_x^{-1}\left(x_0,\hat{\xi}\right)$
$\pi_x^{-1}\left(\hat{x}_0,\xi\right)$
INTERMEDIATE FIBER SUBBUNDLE 1:
THE BOUQUETS $X_t^{\xi}\left(\hat{x}_0\right)$ OF INTEGRAL CURVES STARTING FROM COMMON POINTS
INTERMEDIATE FIBER SUBBUNDLE 2:
THE MOTION SPACES $X_t\left(\hat{\xi}\right)$ AS THE SHEAVES OF INTEGRAL CURVES
$X_t^{\xi}\left(\hat{x}_0'\right)$
$X_t^{\xi}\left(\hat{x}_0''\right)$
$X_t^{\xi}\left(\hat{x}_0'''\right)$
$X_t\left(\hat{\xi}'\right)$
$X_t\left(\hat{\xi}''\right)$
$X_t\left(\hat{\xi}'''\right)$
$G_{\hat{x}_0',\xi}$
$G_{\hat{x}_0'',\xi}$
$G_{\hat{x}_0''',\xi}$
$G_{x_0,\hat{\xi}'}$
$G_{x_0,\hat{\xi}''}$
$G_{x_0,\hat{\xi}'''}$
$X_{t_0}$
$\pi_x^{-1}\left(\hat{x}_0',\xi\right)$
$\pi_x^{-1}\left(\hat{x}_0'',\xi\right)$
$\pi_x^{-1}\left(\hat{x}_0''',\xi\right)$
$\hat{x}_0'$
$\hat{x}_0''$
$\hat{x}_0'''$
$\times$
$\pi_x^{-1}\left(x_0,\hat{\xi}'\right)$
$\pi_x^{-1}\left(x_0,\hat{\xi}''\right)$
$\pi_x^{-1}\left(x_0,\hat{\xi}'''\right)$
$\hat{\xi}'$
$\hat{\xi}''$
$\hat{\xi}'''$
$\pi_x^{-1}\left(\hat{x}_0,\xi\right)$
$X_{t_0}$
$\Xi$
$\pi_x^{-1}\left(x_0,\hat{\xi}\right)$

**FIG. 2**

Suppose the inclusion has a complete set of independent parametric first integrals

$$g(t,x;\xi)=\left(g_1(t,x;\xi),...,g_n(t,x;\xi)\right)\in C^{r+1}\ , \tag{11}$$

which are dependent of the vector of parameters $\xi$.

Set up the following vector equation

$$g(t,x;\xi)=c \Leftrightarrow \left\{g_i(t,x;\xi)=c_i, \forall c_i=g_i(t_0,x_0;\xi)\in R^1, i=(1,...,n)\right\}, \tag{12}$$

where $c=(c_1,...,c_n)=g(t_0,x_0;\xi)=\left(g_1(t_0,x_0;\xi),...,\mathrm{g}_n(t_0,x_0;\xi)\right)$; $(x_0,\xi)\in X_{t_0}\times\Xi$. At some fixed $c=\hat{c}$ and $\xi=\hat{\xi}$ the equation (12) defines the set of $n$ $n$-dimensional invariant manifolds $\left\{M_{x_1}\left(\hat{c}_1,\hat{\xi}\right),....,M_{x_n}\left(\hat{c}_n,\hat{\xi}\right)\right\}$, where

$$M_{x_i}\left(\hat{c}_i,\hat{\xi}\right)=\left\{g_i\left(t,x;\hat{\xi}\right)=\hat{c}_i,(t,x)\in T\times R_x^n;\hat{\xi}\in\Xi\right\}. \tag{13}$$

Thus, in its turn, the following vector equation for $x_0$, which we consider parametric variables,

$$\hat{c}=g\left(t_0,x_0;\hat{\xi}\right) \tag{14}$$

defines the set of $n$ $(n-1)$-dimensional submanifolds $\left\{L_{x_1}\left(\hat{c}_1,\hat{\xi}\right),....,L_{x_n}\left(\hat{c}_n,\hat{\xi}\right)\right\}\subset X_{t_0}$ of the corresponding above-mentioned manifolds $\left\{M_{x_1}\left(\hat{c}_1,\hat{\xi}\right),....,M_{x_n}\left(\hat{c}_n,\hat{\xi}\right)\right\}$ that belong to $T\times R_x^n$. The former ones we obtain by means of the intersection the latter ones with the hyperplane $\{t=t_0\}$ that is

$$\left\{L_{x_1}\left(\hat{c}_1,\hat{\xi}\right),....,L_{x_n}\left(\hat{c}_n,\hat{\xi}\right)\right\}=\left\{M_{x_1}\left(\hat{c}_1,\hat{\xi}\right)\cap\{t=t_0\},....,M_{x_n}\left(\hat{c}_n,\hat{\xi}\right)\cap\{t=t_0\}\right\} \tag{15}$$

where $L_{x_i}\left(\hat{c}_i,\hat{\xi}\right)=\left\{\hat{c}_i=g_i\left(t_0,x_0;\hat{\xi}\right),x_0\in X_{t_0};\hat{\xi}\in\Xi\right\},i=(1,...,n)$.

First, we assume that each $i$-component equation of the vector equation

$$\left\{g_i(t,x;\xi)=c_i\big|_{c_i=g_i(t_0,x_0;\xi)}\right\}_{i=1}^{n}\ , \tag{16}$$

obtained from (12), can be solved for the corresponding component $x_i$ in a unique manner

$$\left\{ g_i\left(t,x;\xi\right)=c_i\Big|_{c_i=g_i\left(t_0,x_0;\xi\right)} \right\}_{i=1}^{n} \Rightarrow \left| \begin{matrix} g_i\left(t,x;\xi\right)=c_i \Rightarrow \\ \Rightarrow x_i=\tilde{\varphi}_i\left(t,x^i;c_i,\xi\right) \end{matrix} \right| \Rightarrow \left\{ x_i=\tilde{\varphi}_i\left(t,x^i;c_i,\xi\right)\Big|_{c_i=g_i\left(t_0,x_0;\xi\right)} \right\}_{i=1}^{n} = \\ =\left\{ x_i=\tilde{\varphi}_i\left(t,x^i;g_i\left(t_0,x_0;\xi\right),\xi\right)\right\}_{i=1}^{n}=\left\{ x_i=\varphi_i\left(t,x^i;t_0,x_0,\xi\right)\right\}_{i=1}^{n} \tag{17}$$

where $x^i=\left(x_1,...,x_{i-1},x_{i+1},...,x_n\right),\varphi_i\left(t,x^i;t_0,x_0,\xi\right)\in C^{r+1},\left(t,x^i\right)\in R^n_{t,x^i},\left(x_0,\xi\right)\in X_{t_0}\times\Xi$ .

In vector form this condition looks as follows

$$x=\varphi\left(t,x;t_0,x_0,\xi\right), \tag{18}$$

where $\varphi\left(t,x;t_0,x_0,\xi\right)=\left(\varphi_1\left(t,x^1;t_0,x_0,\xi\right),...,\varphi_k\left(t,x^k;t_0,x_0,\xi\right),...,\varphi_n\left(t,x^n;t_0,x_0,\xi\right)\right)$,

$k\in\left(2,...,n-1\right),x^1=\left(x_2,...,x_n\right),x^n=\left(x_1,...,x_{n-1}\right)$ .

***Remark 1.*** *The expression from (17) means that the 1-codimensional or $n$-dimensional invariant manifolds $\left\{M_{x_1}\left(\hat{c}_1,\hat{\xi}\right),....,M_{x_n}\left(\hat{c}_n,\hat{\xi}\right)\right\}$ can be equally represented by explicit and implicit functional equations, namely*

$$\left\{ M_{x_i}\left(\hat{c}_i,\hat{\xi}\right)=\left\{ g_i\left(t,x;\hat{\xi}\right)=\hat{c}_i\Big|_{\hat{c}_i=g_i\left(t_0,\hat{x}_0;\hat{\xi}\right)} \Leftrightarrow x_i=\left\{ \begin{matrix} =\tilde{\varphi}_i\left(t,x^i;\hat{c}_i,\hat{\xi}\right) \\ =\varphi_i\left(t,x^i;t_0,\hat{x}_0,\hat{\xi}\right) \end{matrix} \right\},\left(t,x\right)\in T\times R^n_x;\hat{\xi}\in\Xi \right\} \right\}_{i=1}^{n} . \tag{19}$$

*As to the explicit expressions representing the 2-codimensional or $\left(n-1\right)$-dimensional submanifolds $\left\{L_{x_i}\left(\hat{c}_i,\hat{\xi}\right)\right\}_{i=1}^{n}$ of the former ones we have to put $\left(t=t_0,x=x_0\right)$ in the left-hand side of the equations (16) and solve each of them for $x_{i,0}$ in the corresponding order, where $i=\left(1,...,n\right)$, then fix the value of $c$ and $\xi$, that is set $c_i=\hat{c}_i,\xi=\hat{\xi}$ as it is shown below*

$$\left\{ g_i\left(t_0,x_0;\xi\right)=c_i \Rightarrow x_{i,0}=\tilde{\varphi}_i\left(t_0,x_0^i,\xi;c_i\right) \Rightarrow \left| \begin{matrix} c_i=\hat{c}_i \\ \xi=\hat{\xi} \end{matrix} \right| \Rightarrow x_{i,0}=\tilde{\varphi}_i\left(t_0,x_0^i,\hat{\xi};\hat{c}_i\right) \right\}_{i=1}^{n} . \tag{20}$$

*Finally we obtain*

$$\left\{ L_{x_i}\left(\hat{c}_i,\hat{\xi}\right)=\left\{ \hat{c}_i=g_i\left(t_0,x_0;\hat{\xi}\right) \Leftrightarrow x_{i,0}=\tilde{\varphi}_i\left(t_0,x_0^i,\hat{\xi};\hat{c}_i\right),x_0\in X_{t_0};\hat{\xi}\in\Xi \right\}\right\}_{i=1}^{n} . \tag{21}$$

Second, assume that the vector equation consisting of $n$ of the component equation (20) can be solved for $x_0$ in a unique manner, namely

$$\begin{pmatrix} \left\{ g_i\left(t_0, x_0; \xi\right) = c_i \right\}_{i=1}^{n} \\ \Updownarrow \\ g\left(t_0, x_0; \xi\right) = c \end{pmatrix} \Rightarrow x_0 = \psi\left(t_0, c, \xi\right), \tag{22}$$

where the smooth vector function $\psi\left(t_0, c, \xi\right) = \left(\psi_1\left(t_0, c, \xi\right), ..., \psi_n\left(t_0, c, \xi\right)\right)$ defines the diffeomorphism $\psi : R^n \times \Xi \to X_{t_0} \forall t_0 \in T$.

Third, let us introduce a new fiber bundle with the projection based on the relation (22)

$$\psi : R^n \times \Xi \to \left\{ \bigcup_{\forall \hat{\xi} \in \Xi} \left( \bigcup_{\forall \hat{c}_i \in R^1} L_{x_i}\left(\hat{c}_i, \hat{\xi}\right) \right) \right\}_{i=1}^{n} = X_{t_0}. \tag{23}$$

Its total space is $X_{t_0} = \left\{ \bigcup_{\forall \xi \in \Xi} \left( \bigcup_{\forall c_i \in R^1} L_{x_i}\left(c_i, \xi\right) \right) \right\}_{i=1}^{n}$, the base space is $R^n \times \Xi$, the typical fiber is

$$\left\{ L_{x_i}\left(\hat{c}_i, \hat{\xi}\right) = \left\{ \hat{c}_i = g_i\left(t_0, x_0; \hat{\xi}\right) \Rightarrow x_{i,0} = \tilde{\varphi}_i\left(t_0, x_0^i, \hat{\xi}; \hat{c}_i\right), x_0 \in X_{t_0}; \hat{\xi} \in \Xi \right\} \right\}_{i=1}^{n} \tag{24}$$

This means that we can construct *the collection of quotient spaces* $\left\{ X_{t_0} / L_{x_i}\left(\hat{c}_i, \hat{\xi}\right) \right\}_{i=1}^{n}$ for $X_{t_0}$ through the homeomorphism $\psi$ as an equivalence relation acting on $X_{t_0}$. Further in our reasoning if the vector of parameter $\xi$ is fixed and equals some $\hat{\xi}$ it is necessary to replace the total base space $X_{t_0}$ with the total base space $X_{t_0} / L_{x_i}\left(\hat{c}_i, \hat{\xi}\right)$ for each specific component $i$ in the fiber bundles generated by the equation (12).

Consider a fiber bundle determined by the equation (12), namely

$$\left\{ \Gamma_{x_i}\left(t, x; c_i, \xi\right), \pi_{x_i}\left(c_i, \xi\right), R^1 \times \Xi, \mathrm{G}_{\hat{c}_i, \hat{\xi}}^{x_i}\left(t, x\right) \right\} \tag{25}$$

with the total space $\Gamma_{x_i}\left(t, x; c_i, \xi\right) = \left\{ g_i\left(t, x; \xi\right) = c_i, \left(t, x\right) \in T \times R_x^n; \forall \left(c_i, \xi\right) \in R^1 \times \Xi \right\}$, the total projection $\pi_{x_i}\left(c_i, \xi\right) : \Gamma_{x_i}\left(t, x; c_i, \xi\right) \to R^1 \times \Xi$, the total base space $R^1 \times \Xi$, the total typical fiber $\Gamma_{x_i}\left(t, x; \hat{c}_i, \hat{\xi}\right) = M_{x_i}\left(\hat{c}_i, \hat{\xi}\right) = \left\{ g_i\left(t, x; \hat{\xi}\right) = \hat{c}_i, \left(t, x\right) \in T \times R_x^n; \left(\hat{c}_i, \hat{\xi}\right) \in R^1 \times \Xi \right\} = \pi_{x_i}^{-1}\left(\hat{c}_i, \hat{\xi}\right)$ and the $(n-1)$-parameter structure Lie group $G^{x_i}_{\hat{c}_i,\hat{\xi}}(t,x)$ acting on it, where the elements of this group are the integral curves $x_t\left(x_0,\hat{\xi}\right)\in\Gamma_{x_i}\left(t,x;\hat{c}_i,\hat{\xi}\right)=M_{x_i}\left(\hat{c}_i,\hat{\xi}\right)$.

Obviously, the fiber bundle (25) can be considered a fiber metabundle with the typical fiber represented by the total spaces of the two other fiber subbundles, namely

1) the fiber bundle or fiber subbundle in the structure of the fiber metabundle (25)

$$\left\{\Gamma_{x_i}\left(t,x;\hat{x}_0,\xi\right),\pi_{x_i}\left(\hat{x}_0,\xi\right)\xleftrightarrow{\hat{c}_i=g_i\left(t_0,\hat{x}_0;\xi\right)}\pi_{x_i}\left(\hat{c}_i,\xi\right),\Xi,G^{x_i}_{\hat{x}_0,\hat{\xi}}(t,x)\right\} \tag{26}$$

with the total subspace

$$\Gamma_{x_i}\left(t,x;\hat{x}_0,\xi\right)=F^{\xi}_{x_i}\left(\hat{x}_0\right)=\left\{\bigcup_{\forall\hat{\xi}\in\Xi}\left\{x_i=\varphi_i\left(t,x^i;t_0,\hat{x}_0,\hat{\xi}\right)\right\},(t,x)\in T\times R^n_x;\hat{x}_0\in X_{t_0}\right\}$$, the section

projection 2 $\left\{\pi_{x_i}\left(\hat{x}_0,\xi\right)\xleftrightarrow{\hat{c}_i=g_i\left(t_0,\hat{x}_0;\xi\right)}\pi_{x_i}\left(\hat{c}_i,\xi\right)\right\}:\Gamma_{x_i}\left(t,x;\hat{x}_0,\xi\right)\to\Xi$, the base subspace $\Xi$, the typical fiber

$$\Gamma_{x_i}\left(t,x;\hat{x}_0,\hat{\xi}\right)\xleftrightarrow{\hat{c}_i=g_i\left(t_0,\hat{x}_0;\hat{\xi}\right)}M_{x_i}\left(\hat{c}_i,\hat{\xi}\right)=\left\{x_i=\varphi_i\left(t,x^i;t_0,\hat{x}_0,\hat{\xi}\right),(t,x)\in T\times R^n_x;\hat{x}_0\in X_{t_0},\hat{\xi}\in\Xi\right\}=$$

$=\pi^{-1}_{x_i}\left(\hat{x}_0,\hat{\xi}\right)\xleftrightarrow{\hat{c}_i=g_i\left(t_0,\hat{x}_0;\hat{\xi}\right)}\pi^{-1}_{x_i}\left(\hat{c}_i,\hat{\xi}\right)$ and the $(n-1)$-parameter structure group $G^{x_i}_{\hat{x}_0,\hat{\xi}}(t,x)$ acting on it, where the elements of this group are integral curves $x_t\left(x_0,\hat{\xi}\right)\in\Gamma_{x_i}\left(t,x;\hat{c}_i,\hat{\xi}\right)=M_{x_i}\left(\hat{c}_i,\hat{\xi}\right)$.

2) the fiber bundle or fiber subbundle in the structure of the fiber metabundle (25)

$$\left\{\Gamma_{x_i}\left(t,x;x_0,\hat{\xi}\right),\pi_{x_i}\left(c_i,\hat{\xi}\right),R^1,G^{x_i}_{\hat{x}_0,\hat{\xi}}(t,x)\right\} \tag{27}$$

with the total subspace $\Gamma_{x_i}\left(t,x;x_0,\hat{\xi}\right)=F_{x_i}\left(\hat{\xi}\right)=$

$$=\left\{\bigcup_{\forall\hat{c}_i\in R^1}\left\{g_i\left(t,x;\hat{\xi}\right)=\hat{c}_i\Big|_{\hat{c}_i=g_i\left(t_0,x_0;\hat{\xi}\right)}\Leftrightarrow x_i=\varphi_i\left(t,x^i;t_0,x_0,\hat{\xi}\right)\right\},(t,x)\in T\times R^n_x;\hat{\xi}\in\Xi\right\}$$, the section

projection 1 $\pi_{x_i}\left(c_i,\hat{\xi}\right):\Gamma_{x_i}\left(t,x;x_0,\hat{\xi}\right)\to R^1$, the base subspace $R^1$, the typical fiber

$$\Gamma_{x_i}\left(t,x;\hat{x}_0,\hat{\xi}\right)\xleftrightarrow{\hat{c}_i=g_i\left(t_0,\hat{x}_0;\hat{\xi}\right)}M_{x_i}\left(\hat{c}_i,\hat{\xi}\right)=\left\{g_i\left(t,x;\hat{\xi}\right)=\hat{c}_i\Big|_{\hat{c}_i=g_i\left(t_0,\hat{x}_0;\hat{\xi}\right)}\Leftrightarrow x_i=\varphi_i\left(t,x^i;t_0,\hat{x}_0,\hat{\xi}\right),\right.$$

$\left.(t,x)\in T\times R^n_x;\hat{c}_i\in R^1,\hat{\xi}\in\Xi\right\}=\pi^{-1}_{x_i}\left(\hat{c}_i,\hat{\xi}\right)$ and the $(n-1)$-parameter structure group $G^{x_i}_{\hat{x}_0,\hat{\xi}}(t,x)$ acting on it, where the elements of this group are integral curves $x_t\left(x_0,\hat{\xi}\right)\in\Gamma_{x_i}\left(t,x;\hat{c}_i,\hat{\xi}\right)$. In fact this fiber bundle is a 1-codimensional foliation.

***Remark 2.*** *Now we will show why the dimension of the structure Lie group* $\mathrm{G}^{x_i}_{\hat{x}_0,\hat{\xi}}\left(t,x\right)$ *is* $\left(n-1\right)$. *Consider the vector equation*

$$g^i\left(t,x;\hat{\xi}\right)=c^i$$

*for* $x^i=\left(x_1,...,x_{i-1},x_{i+1},...,x_n\right)$, *where* $c^i=\left(c_1,...,c_{i-1},c_{i+1},...,c_n\right)$, $g^i\left(t,x;\hat{\xi}\right)=\left(g_1\left(t,x;\hat{\xi}\right),...,\right.$ $\left.g_{i-1}\left(t,x;\hat{\xi}\right),g_{i+1}\left(t,x;\hat{\xi}\right),...,g_n\left(t,x;\hat{\xi}\right)\right)$. *Now plug* $x_i=\tilde{\varphi}_i\left(t,x^i;\hat{c}_i,\hat{\xi}\right)$ *from (19) in it*

$$g^i\left(t,x^i,\tilde{\varphi}_i\left(t,x^i;\hat{c}_i,\hat{\xi}\right);\hat{\xi}\right)=\tilde{g}^i\left(t,x^i;\hat{c}_i,\hat{\xi}\right)=c^i .$$

*Solve the obtained vector equation for* $x^i$. *We receive*

$$x^i=p^i\left(t;\hat{c}_i,\hat{\xi},c^i\right),$$

*where* $p^i\left(t;\hat{c}_i,\hat{\xi},c^i\right)=\left(p_1\left(t;\hat{c}_i,\hat{\xi},c^i\right),...,p_{i-1}\left(t;\hat{c}_i,\hat{\xi},c^i\right),p_{i+1}\left(t;\hat{c}_i,\hat{\xi},c^i\right),...,p_n\left(t;\hat{c}_i,\hat{\xi},c^i\right)\right)$.

*We shouldn't forget that* $\hat{c}_i,\hat{\xi}$ *are some concrete numbers and in general they can be left out in formulas. The last vector equation describes the orthogonal projection of the family of integral curves given as follows*

$$x=\left(p_1\left(t;\hat{c}_i,\hat{\xi},c^i\right),...,p_{i-1}\left(t;\hat{c}_i,\hat{\xi},c^i\right),\tilde{\varphi}_i\left(t,p^i\left(t;\hat{c}_i,\hat{\xi},c^i\right);\hat{c}_i,\hat{\xi}\right),p_{i+1}\left(t;\hat{c}_i,\hat{\xi},c^i\right),...,p_n\left(t;\hat{c}_i,\hat{\xi},c^i\right)\right)$$

*in* $R^{n-1}_{x^i}$, *which depends on* $\left(n-1\right)$ *parameters* $c^i$ *and compose the* $n$*-dimensional manifold* $\Gamma_{x_i}\left(t,x;\hat{c}_i,\hat{\xi}\right)=M_{x_i}\left(\hat{c}_i,\hat{\xi}\right)$.

***Remark 3.*** *Once again it is very important to emphasize that despite the fact that* $x_0$ *and* $\xi$ *are considered the vectors of parameters, they do not have the equal rights in the hierarchical structure of the fiber metabundle (25) generated by the complete set of the first integrals (11) (or (12)) of the differential inclusion (1). The vector of parameters* $\xi\in\Xi$ *has the first-grade or absolute independence. Meanwhile* $x_0\in X_{t_0}$ *has the second-grade independence since* $c=\left(c_1,...,c_n\right)$ *breaks* $X_{t_0}$ *in the equivalence classes* $L_{x_i}\left(\hat{c}_i,\hat{\xi}\right)\cong\psi_i\left(\hat{c}_i,\hat{\xi}\right)$ *creating the quotient* *space* $X_{t_0} / L_{x_i}\left(\hat{c}_i, \hat{\xi}\right)$ *from* $X_{t_0}$. *It is clearly seen from (22). Fixing* $x_0$ *we choose some concrete point* $\hat{x}_0$ *of phase space and are able to investigate the bouquet of leaves having the point* $\hat{x}_0$ *as common one of different 1-codimensional foliations corresponding to the different points of* $\hat{\xi} \in \Xi$. *Fixing* $\xi$ *we pick out some specific 1-codimensional foliation being the sheaf of 1-codimensional manifolds having the latitude to investigate the properties of its leaves. See Fig.3.*

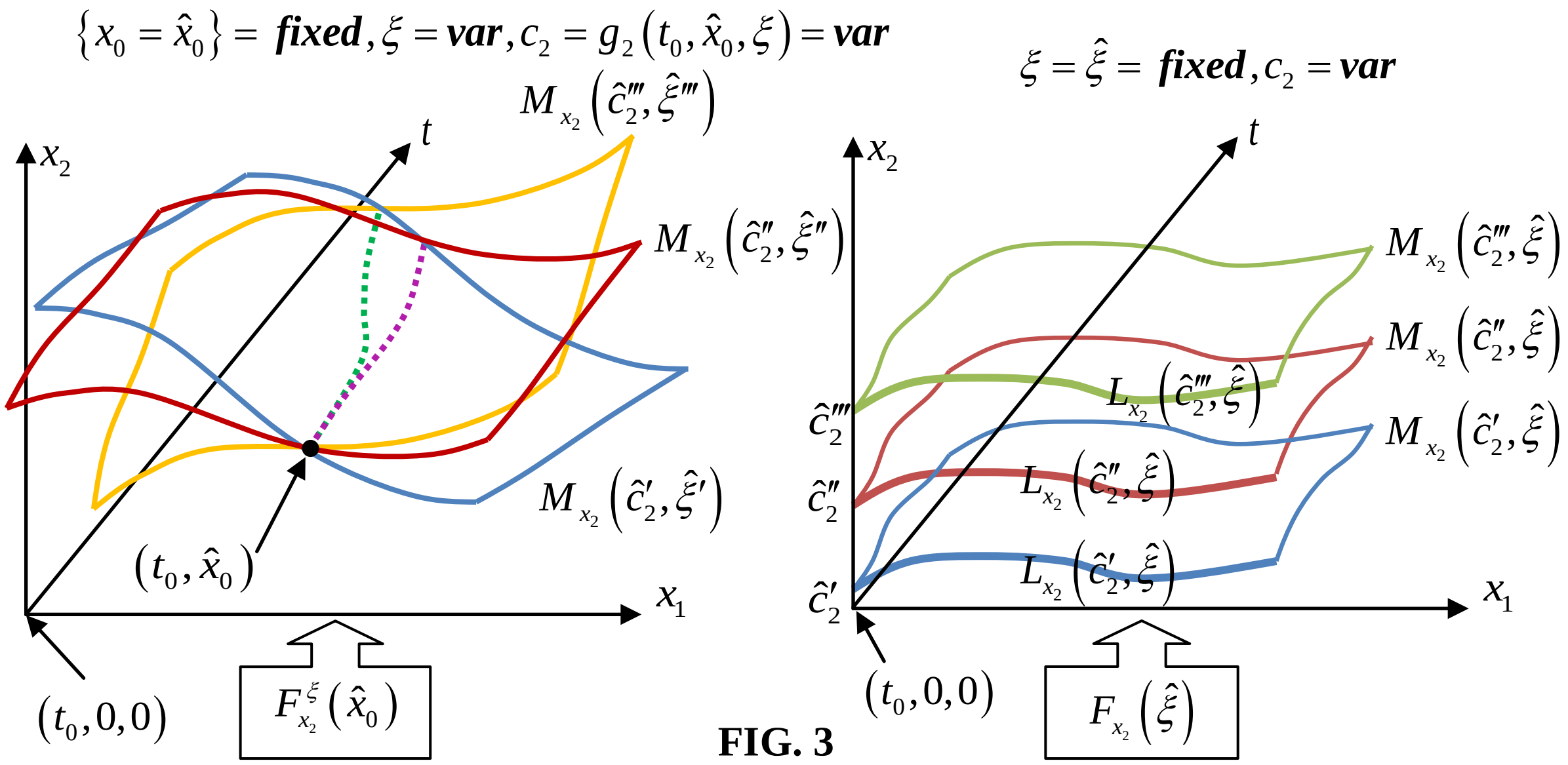

**FIG. 3**

The situation with $x_0$ and $\xi$ reminds the one with commuting and anticommuting variables in supermanifolds [3] and can be investigated additionally if the components of $\xi$ are anticommuting ones.

On the analogy of the fiber bundle (5) we can present the fiber bundle (25) in two ways as follows

1) a fiber metabundle

$$\left\{\left\{\begin{array}{ccc}\Gamma_{x_i}(t,x;x_0,\xi) & \xleftrightarrow{c_i=g_i\left(t_0,x_0;\hat{\xi}\right)} & \Gamma_{x_i}(t,x;c_i,\xi) \\ \pi_{x_i}\left(x_0,\hat{\xi}\right) & \xleftrightarrow{c_i=g_i\left(t_0,x_0;\hat{\xi}\right)} & \pi_{x_i}\left(c_i,\hat{\xi}\right)\end{array}\right\}, X_{t_0}, G^{x_i}_{\hat{x}_0,\xi}(t,x)\right\} \tag{28}$$

with the total metabundle space $\Gamma_{x_i}(t,x;x_0,\xi) \xleftrightarrow{c_i=g_i\left(t_0,x_0;\hat{\xi}\right)} \Gamma_{x_i}(t,x;c_i,\xi) =$

$= \left\{ g_i(t,x;\xi) = c_i\big|_{\hat{c}_i=g_i(t_0,x_0;\xi)} \Leftrightarrow x_i = \varphi_i\left(t,x^i;t_0,x_0,\xi\right),(t,x)\in T\times R^n_x; \forall (x_0,\xi)\in X_{t_0}\times\Xi \right\}$, the section projection 1 $\pi_{x_i}\left(x_0,\hat{\xi}\right) \xleftrightarrow{c_i=g_i\left(t_0,x_0;\hat{\xi}\right)} \pi_{x_i}\left(c_i,\hat{\xi}\right)$, the total base space $X_{t_0}$ and the $m$ -

parameter structure group $G^{x_i}_{\hat{x}_0,\xi}(t,x)$ acting on the typical fiber $\Gamma_{x_i}(t,x;\hat{x}_0,\xi)=F^{\xi}_{x_i}(\hat{x}_0)=$

$=\left\{\bigcup_{\forall\xi\in\Xi}\left\{x_i=\varphi_i\left(t,x^i;t_0,\hat{x}_0,\xi\right)\right\},(t,x)\in T\times R^n_x;\hat{x}_0\in X_{t_0}\right\}=\pi^{-1}_{x_i}(\hat{x}_0,\xi)$ being the total subspace of the fiber bundle (26), where $F^{\xi}_{x_i}(\hat{x}_0)=\bigcup_{\forall\xi\in\Xi}\left\{x_i=\varphi_i\left(t,x^i;t_0,\hat{x}_0,\xi\right)\right\}$ is a bouquet of the invariant manifolds of 1-codimension having the initial point $\hat{x}_{t_0}=(t_0,\hat{x}_0)$ of some specific integral curves $x_t$ as common one.

2) a fiber metabundle

$$\left\{\Gamma_{x_i}(t,x;x_0,\xi)\xleftrightarrow{c_i=g_i\left(t_0,x_0;\hat{\xi}\right)}\Gamma_{x_i}(t,x;c_i,\xi),\pi_{x_i}(\hat{c}_i,\xi),\Xi,G^{x_i}_{x_0,\hat{\xi}}(t,x)\right\} \tag{29}$$

with the total metabundle space $\Gamma_{x_i}(t,x;x_0,\xi)\xleftrightarrow{c_i=g_i\left(t_0,x_0;\hat{\xi}\right)}\Gamma_{x_i}(t,x;c_i,\xi)=$

$=\left\{g_i(t,x;\xi)=c_i\big|_{\hat{c}_i=g_i(t_0,x_0;\xi)}\Leftrightarrow x_i=\varphi_i\left(t,x^i;t_0,x_0,\xi\right),(t,x)\in T\times R^n_x;\forall(x_0,\xi)\in X_{t_0}\times\Xi\right\}$, the section projection 2 $\pi_{x_i}(\hat{c}_i,\xi)$, the total base space $\Xi$, and the $n$-parameter structure group $G^{x_i}_{x_0,\hat{\xi}}(t,x)$ that acts on the typical fiber $\Gamma_{x_i}\left(t,x;c_i,\hat{\xi}\right)=\Gamma_{x_i}\left(t,x;x_0,\hat{\xi}\right)=F_{x_i}\left(\hat{\xi}\right)=$

$=\left\{\bigcup_{\forall c_i\in R^1}\left\{g_i\left(t,x;\hat{\xi}\right)=c_i\big|_{c_i=g_i\left(t_0,x_0;\hat{\xi}\right)}\Leftrightarrow x_i=\varphi_i\left(t,x^i;t_0,x_0,\hat{\xi}\right)\right\},(t,x)\in T\times R^n_x;\hat{\xi}\in\Xi\right\}=\pi^{-1}_{x_i}\left(c_i,\hat{\xi}\right)$,

which is in its turn the total subspace 2 of the fiber bundle (27). We suppose here that $F_{x_i}\left(\hat{\xi}\right)=\bigcup_{\forall c_i\in R^1}\left\{g_i\left(t,x;\hat{\xi}\right)=c_i\big|_{c_i=g_i\left(t_0,x_0;\hat{\xi}\right)}\right\}$ is a foliation of the invariant manifolds of 1-codimension being the union of all the sections of the first integral $g_i(t,x;\xi)$ created by hyperplanes $c_i=\hat{c}_i\ \forall\hat{c}_i\in R^1$ at some fixed point $\xi=\hat{\xi}\in\Xi$.

## 2. PROCEDURE OF FLATTENING TOTAL SPACES OF FIBER BUNDLES

### *The initial system and the cascade of flattening diffeomorphisms*

Now once again we indicate the whole initial system that will be subject to the procedure of sequential transformations by means of the cascade of flattening diffeomorphisms. The one is given below

$$\left\{ \left\{ \Gamma_{x_i}\left(t,x;\mathrm{c}_i,\xi\right), \pi_{x_i}\left(\mathrm{c}_i,\xi\right), R^1\times\Xi, \mathrm{G}^{x_i}_{\hat{c}_i,\hat{\xi}}\left(t,x\right) \right\}_{i=1}^{n}; \frac{dx}{dt}\in\bigcup_{\xi\in\Xi} f\left(t,x;\xi\right) \right\}, \tag{30}$$

where $f\left(t,x;\xi\right)=\left(f_1\left(t,x;\xi\right),...,f_n\left(t,x;\xi\right)\right)\in C^r, g\left(t,x;\xi\right)=C^{r+1}$. As it can be seen from (30) the whole initial system is composed of the differential inclusion (1) and the set of all the fiber metabundles (25) for all the components of the phase vector $x$.

The cascade of sequential flattening diffeomorphisms is defined as

$$\theta=\left\{\begin{array}{l}
\underbrace{\theta_1=\left\{x_1=y_1+\varphi_1\left(t,x_2,...,x_n;c_1,\xi\right)\right\}:\Gamma_{x_1}\left(t,x;c_1,\xi\right)\to\left\{y_1=0\right\}\subset R^{2n+m+1}}\\
\Downarrow\\
\underbrace{\theta_2=\left\{x_2=y_2+\varphi_2^1\left(t,y_1,x_3,...,x_n;c_1,c_2,\xi\right)\right\}:\Gamma_{x_2}\left(t,x;c_2,\xi\right)\to\left\{y_2=0\right\}\subset R^{2n+m+1}}\\
\Downarrow\\
\theta_3=\left\{x_3=y_3+\varphi_3^2\left(t,y_1,y_2,x_4,...,x_n;c_1,c_2,c_3,\xi\right)\right\}:\Gamma_{x_3}\left(t,x;c_3,\xi\right)\to\\
\underbrace{\to\left\{y_3=0\right\}\subset R^{2n+m+1}}\\
\Downarrow\\
\dots\dots\dots\dots\dots\dots\dots\dots\dots\dots\dots\dots\dots\dots\dots\dots\\
\theta_{k-1}=\left\{x_{k-1}=y_{k-1}+\varphi_{k-1}^{k-2}\left(t,y_1,...,y_{k-2},x_k,...,x_n;c_1,...,c_{k-1},\xi\right)\right\}:\Gamma_{x_{k-1}}\left(t,x;c_{k-1},\xi\right)\to\\
\underbrace{\to\left\{y_{k-1}=0\right\}\subset R^{2n+m+1}}\\
\Downarrow\\
\theta_k=\left\{x_k=y_k+\varphi_k^{k-1}\left(t,y_1,...,y_{k-1},x_{k+1},...,x_n;c_1,...,c_k,\xi\right)\right\}:\Gamma_{x_k}\left(t,x;c_k,\xi\right)\to\\
\underbrace{\to\left\{y_k=0\right\}\subset R^{2n+m+1}}\\
\Downarrow\\
\dots\dots\dots\dots\dots\dots\dots\dots\dots\dots\dots\dots\dots\dots\dots\dots\\
\theta_{n-1}=\left\{x_{n-1}=y_{n-1}+\varphi_{n-1}^{n-2}\left(t,y_1,...,y_{n-2},x_n;c_1,...,c_{n-1},\xi\right)\right\}:\Gamma_{x_{n-1}}\left(t,x;c_{n-1},\xi\right)\to\\
\underbrace{\to\left\{y_{n-1}=0\right\}\subset R^{2n+m+1}}\\
\Downarrow\\
\theta_n=\left\{x_n=y_n+\varphi_n^{n-1}\left(t,y_1,...,y_{n-1};c_1,...,c_n,\xi\right)\right\}:\Gamma_{x_n}\left(t,x;c_n,\xi\right)\to\\
\to\left\{y_n=0\right\}\subset R^{2n+m+1}
\end{array}\right\}, \tag{31}$$

where $\theta=\left(\theta_1\Rightarrow\theta_2\Rightarrow\theta_3\Rightarrow...\Rightarrow\theta_{k-1}\Rightarrow\theta_k\Rightarrow...\Rightarrow\theta_{n-1}\Rightarrow\theta_n\right)$.

In fact, the flattening diffeomorphisms are the orthogonal projections of geometrical-topological objects embedded namely, $\left\{\Gamma_{x_i}\left(t,x;\mathrm{c}_i,\xi\right)\right\}_{i=1}^{n}$ in $R^{2n+m+1}_{t,x,c,\xi}$ onto its $2n+m$-dimensional subspaces

$\left\{R^{2n+m}_{t,x^1,c,\xi}, R^{2n+m}_{t,x^2,c,\xi}, R^{2n+m}_{t,x^3,c,\xi},..., R^{2n+m}_{t,x^{k-1},c,\xi}, R^{2n+m}_{t,x^k,c,\xi},..., R^{2n+m}_{t,x^{n-1},c,\xi}, R^{2n+m}_{t,x^n,c,\xi}\right\}$.

***The first stage of the cascade***

The flattening diffeomorphisms $\theta_1$ transforms

1) the set of the fiber metabundles from (30) into the set of following ones

$$\left\{\Gamma^1_{x_i}\left(t,y_1,x_2,...,x_n;c_1,c_i,\xi\right),\pi^1_{x_i}\left(c_1,\xi\right),R^1\times\Xi,\mathrm{G}^{x_i}_{\hat{c},\hat{\xi}}\left(t,y_1,x_2,...,x_n\right)^1\right\}_{i=1}^{n} \qquad (32)$$

with the total spaces being

$$\left\{\begin{array}{l}\Gamma^1_{x_1}\left(t,y_1,x_2,...,x_n;c_1,c_i,\xi\right)=\left\{y_1=0\right\},\\ \left\{\Gamma^1_{x_i}\left(t,y_1,x_2,...,x_n;c_1,c_i,\xi\right)=\left(\begin{array}{l}g_i\left(t,y_1+\varphi_1\left(t,x_2,...,x_n;c_1,\xi\right),x_2,...,x_n,\xi\right)=\\ =g_i^1\left(t,y_1,x_2,...,x_n;c_1,\xi\right)=c_i\end{array}\right)\right\}_{i=2}^{n}\end{array}\right\}. \qquad (33)$$

Here we assume that

$$g_1\left(t,y_1+\varphi_1\left(t,x_2,...,x_n;c_1,\xi\right),x_2,...,x_n,\xi\right)-c_1\equiv 0\Big|_{y_1=0}\ ; \qquad (34)$$

2) the differential inclusion from (30) into the differential inclusion

$$\left\{\begin{array}{l}\dfrac{dy_1}{dt}\in\bigcup\limits_{\xi\in\Xi}f_1^1\left(t,y_1,x_2,...,x_n;c_1,\xi\right),\dfrac{dx_2}{dt}\in\bigcup\limits_{\xi\in\Xi}f_2^1\left(t,y_1,x_2,...,x_n;c_1,\xi\right),...\\ ...,\dfrac{dx_n}{dt}\in\bigcup\limits_{\xi\in\Xi}f_n^1\left(t,y_1,x_2,...,x_n;c_1,\xi\right)\end{array}\right\}, \qquad (35)$$

where $f_1^1\left(t,y_1,x_2,...,x_n;c_1,\xi\right)=f_1\left(t,y_1+\varphi_1\left(t,x_2,...,x_n;c_1,\xi\right),x_2,...,x_n,\xi\right)-\dfrac{\partial\varphi_1\left(t,x_2,...,x_n;c_1,\xi\right)}{\partial t}-$

$-\sum\limits_{i=2}^{n}\dfrac{\partial\varphi_1\left(t,x_2,...,x_n;c_1,\xi\right)}{\partial x_i}f_i\left(t,y_1+\varphi_1\left(t,x_2,...,x_n;c_1,\xi\right),x_2,...,x_n,\xi\right)$, $f_i^1\left(t,y_1,x_2,...,x_n;c_1,\xi\right)=$

$=f_i\left(t,y_1+\varphi_1\left(t,x_2,...,x_n;c_1,\xi\right),x_2,...,x_n,\xi\right),i=\left(2,...,n\right)$.

The $X_{t_0}$ transforms into $X_{t_0}^{y_1}=\theta_1\left(X_{t_0}\right)$ that is the orthogonal projection of $X_{t_0}$ onto $R^{2n+m}_{t,x^1,c,\xi}$, that is $X_{t_0}^{y_1}\subseteq R^{2n+m}_{t,x^1,c,\xi}$.

Thus at the end of the first stage of the cascade of flattening diffeomorphisms we receive the following whole after-first-stage-of-cascade-of-diffeomorphisms system

$$\left\{\begin{array}{l}\left\{\Gamma^1_{x_i}\left(t,y_1,x_2,...,x_n;c_i,\xi\right),\pi^1_{x_i}\left(c_i,\xi\right),R^1\times\Xi,\mathrm{G}^{x_i}_{\hat{c}_i,\hat{\xi}}\left(t,y_1,x_2,...,x_n\right)^1\right\}_{i=1}^{n};\\ \left\{\begin{array}{l}\frac{dy_1}{dt}\in\bigcup_{\xi\in\Xi}f_1^1\left(t,y_1,x_2,...,x_n;c_1,\xi\right),\frac{dx_2}{dt}\in\bigcup_{\xi\in\Xi}f_2^1\left(t,y_1,x_2,...,x_n;c_1,\xi\right),...\\ ...,\frac{dx_n}{dt}\in\bigcup_{\xi\in\Xi}f_n^1\left(t,y_1,x_2,...,x_n;c_1,\xi\right)\end{array}\right\}\end{array}\right\}, \tag{36}$$

where $f_1^1\left(t,y_1,x_2,...,x_n;c_1,\xi\right)\Big|_{\forall\left(t,y_1,x_2,...,x_n;c,\xi\right)\in\Gamma^1_{x_1}\left(t,y_1,x_2,...,x_n;c_1,\xi\right)=\{y_1=0\}}\equiv 0$, $y_{1,0}=0$.

### *The second stage of the cascade*

The flattening diffeomorphisms $\theta_2$ transforms

1) the set of the fiber metabundles from (36) into the set of following ones

$$\left\{\Gamma^2_{x_i}\left(t,y_1,y_2,x_3,...,x_n;c_1,c_2,c_i,\xi\right),\pi^2_{x_i}\left(c_1,c_2,\xi\right),R^1\times\Xi,\mathrm{G}^{x_i}_{\hat{c},\hat{\xi}}\left(t,y_1,y_2,x_3,...,x_n\right)^2\right\}_{i=1}^{n} \tag{37}$$

with the total spaces being

$$\left\{\begin{array}{l}\Gamma^2_{x_1}\left(t,y_1,y_2,x_3,...,x_n;c_1,c_2,c_i,\xi\right)=\left\{y_1=0\right\},\Gamma^2_{x_2}\left(t,y_1,y_2,x_3,...,x_n;c_1,c_2,c_i,\xi\right)=\left\{y_2=0\right\},\\ \left\{\begin{array}{l}\Gamma^2_{x_i}\left(t,y_1,x_2,...,x_n;c_1,c_2,c_i,\xi\right)=\\ =\left(\begin{array}{l}g_i^1\left(t,y_1,y_2+\varphi_2^1\left(t,y_1,x_3,...,x_n;c_1,c_2,\xi\right),x_3,...,x_n;c_1,\xi\right)=\\ =g_i^2\left(t,y_1,y_2,x_3,...,x_n;c_1,c_2,\xi\right)=c_i\end{array}\right)\end{array}\right\}_{i=3}^{n}\end{array}\right\}. \tag{38}$$

Here we assume that

$$g_2^1\left(t,y_1,y_2+\varphi_2^1\left(t,y_1,x_3,...,x_n;c_1,c_2,\xi\right),x_3,...,x_n,\xi\right)-c_2\equiv 0\Big|_{y_2=0}; \tag{39}$$

2) the differential inclusion from (36) into the differential inclusion

$$\left\{\begin{array}{l}\frac{dy_1}{dt}\in\bigcup_{\xi\in\Xi}f_1^2\left(t,y_1,y_2,x_3,...,x_n;c_1,c_2,\xi\right),\frac{dy_2}{dt}\in\bigcup_{\xi\in\Xi}f_2^2\left(t,y_1,y_2,x_3,...,x_n;c_1,c_2,\xi\right),\\ \frac{dx_3}{dt}\in\bigcup_{\xi\in\Xi}f_3^2\left(t,y_1,y_2,x_3,...,x_n;c_1,c_2,\xi\right),...,\frac{dx_n}{dt}\in\bigcup_{\xi\in\Xi}f_n^2\left(t,y_1,y_2,x_3,...,x_n;c_1,c_2,\xi\right)\end{array}\right\}, \tag{40}$$

where $f_1^2\left(t,y_1,y_2,x_3,...,x_n;c_1,c_2,\xi\right)=f_1^1\left(t,y_1,y_2+\varphi_2^1\left(t,y_1,x_3,...,x_n;c_1,c_2,\xi\right),x_3,...,x_n;c_1,\xi\right)$,

$$f_2^2\left(t,y_1,y_2,x_3,...,x_n;c_1,c_2,\xi\right)=f_2^1\left(t,y_1,y_2+\varphi_2^1\left(t,y_1,x_3,...,x_n;c_1,c_2,\xi\right),x_3,...,x_n;c_1,\xi\right)-$$
$$-\frac{\partial\varphi_2^1\left(t,y_1,x_3,...,x_n;c_1,c_2,\xi\right)}{\partial t}-\frac{\partial\varphi_2^1\left(t,y_1,x_3,...,x_n;c_1,c_2,\xi\right)}{\partial y_1}f_1^2\left(t,y_1,y_2,x_3,...,x_n;c_1,c_2,\xi\right)-$$
$$-\sum_{i=3}^{n}\frac{\partial\varphi_2^1\left(t,y_1,x_3,...,x_n;c_1,c_2,\xi\right)}{\partial x_i}f_i^2\left(t,y_1,y_2,x_3,...,x_n;c_1,c_2,\xi\right),\ f_3^2\left(t,y_1,y_2,x_3,...,x_n;c_1,c_2,\xi\right)=$$
$$=f_3^1\left(t,y_1,y_2+\varphi_2^1\left(t,y_1,x_3,...,x_n;c_1,c_2,\xi\right),x_3,...,x_n;c_1,\xi\right),...,\ f_n^2\left(t,y_1,y_2,x_3,...,x_n;c_1,c_2,\xi\right)=$$
$$=f_n^1\left(t,y_1,y_2+\varphi_2^1\left(t,y_1,x_3,...,x_n;c_1,c_2,\xi\right),x_3,...,x_n;c_1,\xi\right).$$

The $X_{t_0}^{y_1}$ transforms into $X_{t_0}^{y_1y_2}=\theta_2\left(X_{t_0}^{y_1}\right)$ that is the orthogonal projection of $X_{t_0}^{y_1}$ onto $R^{2n+m}_{t,y_1,x_3,...,x_n,c,\xi}$, that is $X_{t_0}^{y_1y_2}\subseteq R^{2n+m}_{t,y_1,x_3,...,x_n,c,\xi}$.

Thus at the end of the second stage of the cascade of flattening diffeomorphisms we receive the following whole after-second-stage-of-cascade-of-diffeomorphisms system

$$\left\{\begin{array}{l}\left\{\Gamma_{x_i}^2\left(t,y_1,y_2,x_3,...,x_n;c_1,c_2,c_i,\xi\right),\pi_{x_i}^2\left(c_1,c_2,\xi\right),R^1\times\Xi,\mathrm{G}_{\hat{c},\hat{\xi}}^{x_i}\left(t,y_1,y_2,x_3,...,x_n\right)^2\right\}_{i=1}^{n};\\ \left\{\begin{array}{l}\dfrac{dy_1}{dt}\in\bigcup\limits_{\xi\in\Xi}f_1^2\left(t,y_1,y_2,x_3,...,x_n;c_1,c_2,\xi\right),\dfrac{dy_2}{dt}\in\bigcup\limits_{\xi\in\Xi}f_2^2\left(t,y_1,y_2,x_3,...,x_n;c_1,c_2,\xi\right),\\ \dfrac{dx_3}{dt}\in\bigcup\limits_{\xi\in\Xi}f_3^2\left(t,y_1,y_2,x_3,...,x_n;c_1,c_2,\xi\right),...,\dfrac{dx_n}{dt}\in\bigcup\limits_{\xi\in\Xi}f_n^2\left(t,y_1,y_2,x_3,...,x_n;c_1,c_2,\xi\right)\end{array}\right\}\end{array}\right\},\quad(41)$$

where $f_1^2\left(t,y_1,y_2,x_3,...,x_n;c_1,c_2,\xi\right)\Big|_{\forall\left(t,y_1,y_2,x_3,...,x_n;c,\xi\right)\in\Gamma_{x_1}^2\left(t,y_1,y_2,x_3,...,x_n;c_1,c_2,c_i,\xi\right)=\{y_1=0\}}\equiv0,y_{1,0}=0$;

$f_2^2\left(t,y_1,y_2,x_3,...,x_n;c_1,c_2,\xi\right)\Big|_{\forall\left(t,y_1,y_2,x_3,...,x_n;c,\xi\right)\in\Gamma_{x_2}^2\left(t,y_1,y_2,x_3,...,x_n;c_1,c_2,c_i,\xi\right)=\{y_2=0\}}\equiv0,y_{2,0}=0$.

### *The third stage of the cascade*

The flattening diffeomorphisms $\theta_3$ transforms

1) the set of the fiber metabundles from (41) into the set of following ones

$$\left\{\Gamma_{x_i}^3\left(t,y_1,y_2,y_3,x_4,...,x_n;c_1,c_2,c_3,c_i,\xi\right),\pi_{x_i}^3\left(c_1,c_2,c_3,\xi\right),R^1\times\Xi,\mathrm{G}_{\hat{c},\hat{\xi}}^{x_i}\left(t,y_1,y_2,y_3,x_4,...,x_n\right)^3\right\}_{i=1}^{n}\quad(42)$$

with the total spaces being

$$\left\{\begin{array}{l}\Gamma^3_{x_1}\left(t,y_1,y_2,y_3,x_4,...,x_n;c_1,c_2,c_3,c_i,\xi\right)=\left\{y_1=0\right\},\\ \Gamma^3_{x_2}\left(t,y_1,y_2,y_3,x_4,...,x_n;c_1,c_2,c_3,c_i,\xi\right)=\left\{y_2=0\right\},\\ \Gamma^3_{x_3}\left(t,y_1,y_2,y_3,x_4,...,x_n;c_1,c_2,c_3,c_i,\xi\right)=\left\{y_3=0\right\},\\ \left\{\begin{array}{l}\Gamma^3_{x_i}\left(t,y_1,y_2,y_3,x_4,...,x_n;c_1,c_2,c_3,c_i,\xi\right)=\\ =\left(\begin{array}{l}g_i^2\left(t,y_1,y_2,y_3+\varphi_3^2\left(t,y_1,y_2,x_4,...,x_n;c_1,c_2,c_3,\xi\right),x_4,...,x_n;c_1,c_2,\xi\right)=\\ =g_i^3\left(t,y_1,y_2,y_3,x_4,...,x_n;c_1,c_2,c_3,\xi\right)=c_i\end{array}\right)\end{array}\right\}_{i=4}^{n}\end{array}\right\}. \tag{43}$$

Here again we assume that

$$g_3^2\left(t,y_1,y_2,y_3+\varphi_3^2\left(t,y_1,y_2,x_4,...,x_n;c_1,c_2,c_3,\xi\right),x_4,...,x_n;c_1,c_{2,}\xi\right)-c_3\equiv 0\Big|_{y_3=0}; \tag{44}$$

2) the differential inclusion from (41) into the differential inclusion

$$\left\{\begin{array}{l}\dfrac{dy_1}{dt}\in\bigcup\limits_{\xi\in\Xi}f_1^3\left(t,y_1,y_2,y_3,x_4,...,x_n;c_1,c_2,c_3,\xi\right),\\ \dfrac{dy_2}{dt}\in\bigcup\limits_{\xi\in\Xi}f_2^3\left(t,y_1,y_2,y_3,x_4,...,x_n;c_1,c_2,c_3,\xi\right),\\ \dfrac{dy_3}{dt}\in\bigcup\limits_{\xi\in\Xi}f_3^3\left(t,y_1,y_2,y_3,x_4,...,x_n;c_1,c_2,c_3,\xi\right),\\ \dfrac{dx_4}{dt}\in\bigcup\limits_{\xi\in\Xi}f_4^3\left(t,y_1,y_2,y_3,x_4,...,x_n;c_1,c_2,c_3,\xi\right),\\ \cdots\cdots\cdots\cdots\cdots\cdots\cdots\cdots\cdots\cdots\\ \dfrac{dx_n}{dt}\in\bigcup\limits_{\xi\in\Xi}f_n^3\left(t,y_1,y_2,y_3,x_4,...,x_n;c_1,c_2,c_3,\xi\right)\end{array}\right\}, \tag{45}$$

where

$$f_1^3\left(t,y_1,y_2,y_3,x_4,...,x_n;c_1,c_2,c_3,\xi\right)=$$
$$=f_1^2\left(t,y_1,y_2,y_3+\varphi_3^2\left(t,y_1,y_2,x_4,...,x_n;c_1,c_2,c_3,\xi\right),x_4,...,x_n;c_1,c_{2,}\xi\right),$$
$$f_2^3\left(t,y_1,y_2,y_3,x_4,...,x_n;c_1,c_2,c_3,\xi\right)=$$
$$=f_2^2\left(t,y_1,y_2,y_3+\varphi_3^2\left(t,y_1,y_2,x_4,...,x_n;c_1,c_2,c_3,\xi\right),x_4,...,x_n;c_1,c_{2,}\xi\right),$$

$$f_3^3(t,y_1,y_2,y_3,x_4,...,x_n;c_1,c_2,c_3,\xi)=$$
$$=f_3^2\left(t,y_1,y_2,y_3+\varphi_3^2(t,y_1,y_2,x_4,...,x_n;c_1,c_2,c_3,\xi),x_4,...,x_n;c_1,c_{2,}\xi\right)-$$
$$-\frac{\partial\varphi_3^2(t,y_1,y_2,x_4,...,x_n;c_1,c_2,c_3,\xi)}{\partial t}-\frac{\partial\varphi_3^2(t,y_1,y_2,x_4,...,x_n;c_1,c_2,c_3,\xi)}{\partial y_1}\times$$
$$\times f_1^3(t,y_1,y_2,y_3,x_4,...,x_n;c_1,c_2,c_3,\xi)-\frac{\partial\varphi_3^2(t,y_1,y_2,x_4,...,x_n;c_1,c_2,c_3,\xi)}{\partial y_2}\times$$
$$\times f_2^3(t,y_1,y_2,y_3,x_4,...,x_n;c_1,c_2,c_3,\xi)-\sum_{i=4}^{n}\frac{\partial\varphi_3^2(t,y_1,y_2,x_4,...,x_n;c_1,c_2,c_3,\xi)}{\partial x_i}\times$$
$$\times f_i^3(t,y_1,y_2,y_3,x_4,...,x_n;c_1,c_2,c_3,\xi),\ f_4^3(t,y_1,y_2,y_3,x_4,...,x_n;c_1,c_2,c_3,\xi)=$$
$$=f_4^2\left(t,y_1,y_2,y_3+\varphi_3^2(t,y_1,y_2,x_4,...,x_n;c_1,c_2,c_3,\xi),x_4,...,x_n;c_1,c_{2,}\xi\right),...,$$
$$f_n^3(t,y_1,y_2,y_3,x_4,...,x_n;c_1,c_2,c_3,\xi)=$$
$$=f_n^3\left(t,y_1,y_2,y_3+\varphi_3^2(t,y_1,y_2,x_4,...,x_n;c_1,c_2,c_3,\xi),x_4,...,x_n;c_1,c_{2,}\xi\right).$$

The $X_{t_0}^{y_1y_2}$ transforms into $X_{t_0}^{y_1y_2y_3}=\theta_3\left(X_{t_0}^{y_1y_2}\right)$ that is the orthogonal projection of $X_{t_0}^{y_1y_2}$ onto $R^{2n+m}_{t,y_1,y_2,x_4,...,x_n,c,\xi}$, that is $X_{t_0}^{y_1y_2y_3}\subseteq R^{2n+m}_{t,y_1,y_2,x_4,...,x_n,c,\xi}$.

Thus at the end of the third stage of the cascade of flattening diffeomorphisms we receive the following whole after-third-stage-of-cascade-of-diffeomorphisms system

$$\left\{\begin{array}{l}\left\{\begin{array}{l}\Gamma_{x_i}^3(t,y_1,y_2,y_3,x_4,...,x_n;c_1,c_2,c_3,c_i,\xi),\pi_{x_i}^3(c_1,c_2,c_3,\xi),R^1\times\Xi,\\ G_{\hat{c},\hat{\xi}}^{x_i}(t,y_1,y_2,y_3,x_4,...,x_n)^3\end{array}\right\}_{i=1}^{n};\\ \left\{\begin{array}{l}\frac{dy_1}{dt}\in\bigcup_{\xi\in\Xi}f_1^3(t,y_1,y_2,y_3,x_4,...,x_n;c_1,c_2,c_3,\xi),\\ \frac{dy_2}{dt}\in\bigcup_{\xi\in\Xi}f_2^3(t,y_1,y_2,y_3,x_4,...,x_n;c_1,c_2,c_3,\xi),\\ \frac{dy_3}{dt}\in\bigcup_{\xi\in\Xi}f_3^3(t,y_1,y_2,y_3,x_4,...,x_n;c_1,c_2,c_3,\xi),\\ \frac{dx_4}{dt}\in\bigcup_{\xi\in\Xi}f_4^3(t,y_1,y_2,y_3,x_4,...,x_n;c_1,c_2,c_3,\xi),\\ \ldots\ldots\ldots\ldots\ldots\ldots\ldots\ldots\ldots\ldots\\ \frac{dx_n}{dt}\in\bigcup_{\xi\in\Xi}f_n^3(t,y_1,y_2,y_3,x_4,...,x_n;c_1,c_2,c_3,\xi)\end{array}\right\}\end{array}\right\},\qquad(46)$$

where

$$f_1^3\left(t,y_1,y_2,y_3,x_4,...,x_n;c_1,c_2,c_3,\xi\right)\Big|_{\forall(t,y_1,y_2,y_3,x_4,...,x_n;c,\xi)\in\Gamma_{x_1}^3(t,y_1,y_2,y_3,x_3,...,x_n;c_1,c_2,c_3,c_i,\xi)=\{y_1=0\}}\equiv 0, y_{1,0}=0;$$

$$f_2^3\left(t,y_1,y_2,y_3,x_4,...,x_n;c_1,c_2,c_3,\xi\right)\Big|_{\forall(t,y_1,y_2,y_3,x_4,...,x_n;c,\xi)\in\Gamma_{x_2}^3(t,y_1,y_2,y_3,x_3,...,x_n;c_1,c_2,c_3,c_i,\xi)=\{y_2=0\}}\equiv 0, y_{2,0}=0;$$

$$f_3^3\left(t,y_1,y_2,y_3,x_4,...,x_n;c_1,c_2,c_3,\xi\right)\Big|_{\forall(t,y_1,y_2,y_3,x_4,...,x_n;c,\xi)\in\Gamma_{x_3}^3(t,y_1,y_2,y_3,x_3,...,x_n;c_1,c_2,c_3,c_i,\xi)=\{y_3=0\}}\equiv 0, y_{3,0}=0.$$

### *The ($k$)-th stage of the cascade*

From the whole after-($k-1$)th-stage-of-cascade-of-diffeomorphisms system the flattening diffeomorphisms $\theta_k$ transforms

1) the set of its fiber metabundles into the set of following ones

$$\left\{\begin{matrix}\Gamma_{x_i}^k\left(t,y_1,...,y_k,x_{k+1},...,x_n;c_1,...,c_k,c_i,\xi\right),\pi_{x_i}^k\left(c_1,...,c_k,\xi\right),R^1\times\Xi,\\ \mathrm{G}_{\hat{c},\hat{\xi}}^{x_i}\left(t,y_1,...,y_k,x_{k+1},...,x_n\right)^k\end{matrix}\right\}_{i=1}^n \tag{47}$$

with the total spaces being

$$\left\{\begin{matrix}\Gamma_{x_1}^k\left(t,y_1,...,y_k,x_{k+1},...,x_n;c_1,...,c_k,c_i,\xi\right)=\left\{y_1=0\right\},\\ \dots\dots\dots\dots\dots\dots\dots\dots\dots\dots\dots\dots\\ \Gamma_{x_k}^k\left(t,y_1,...,y_k,x_{k+1},...,x_n;c_1,...,c_k,c_i,\xi\right)=\left\{y_k=0\right\},\\ \left\{\begin{matrix}\Gamma_{x_i}^k\left(t,y_1,...,y_k,x_{k+1},...,x_n;c_1,...,c_k,c_i,\xi\right)=\\ =\left(\begin{matrix}g_i^{k-1}(t,y_1,...,y_{k-1},y_k+\varphi_k^{k-1}\left(t,y_1,...,y_{k-1},x_{k+1},...,x_n;c_1,...,c_k,\xi\right),x_{k+1},...,\\ x_n;c_1,...,c_{k-1,}\xi)=g_i^k\left(t,y_1,...,y_k,x_{k+1},...,x_n;c_1,...,c_k,\xi\right)=c_i\end{matrix}\right)\end{matrix}\right\}_{i=k+1}^n\end{matrix}\right\}. \tag{48}$$

Here we assume that

$$\left.\begin{matrix}g_k^{k-1}\left(t,y_1,...,y_{k-1},y_k+\varphi_k^{k-1}\left(t,y_1,...,y_{k-1},x_{k+1},...,x_n;c_1,...,c_k,\xi\right),x_{k+1},...,x_n;c_1,...,c_{k-1,}\xi\right)-\\ -c_k\equiv 0\end{matrix}\right|_{y_k=0}; \tag{49}$$

2) its differential inclusion into the differential inclusion

$$\left\{\begin{array}{l} \dfrac{dy_1}{dt} \in \bigcup_{\xi\in\Xi} f_1^k\left(t, y_1,..., y_k, x_{k+1},..., x_n; c_1,..., c_k, \xi\right), \\ \text{.............................................................} \\ \dfrac{dy_3}{dt} \in \bigcup_{\xi\in\Xi} f_k^k\left(t, y_1,..., y_k, x_{k+1},..., x_n; c_1,..., c_k, \xi\right), \\ \dfrac{dx_4}{dt} \in \bigcup_{\xi\in\Xi} f_{k+1}^k\left(t, y_1,..., y_k, x_{k+1},..., x_n; c_1,..., c_k, \xi\right), \\ \text{.............................................................} \\ \dfrac{dx_n}{dt} \in \bigcup_{\xi\in\Xi} f_n^k\left(t, y_1,..., y_k, x_{k+1},..., x_n; c_1,..., c_k, \xi\right) \end{array}\right\}, \quad (50)$$

where

$$f_1^k\left(t, y_1,..., y_k, x_{k+1},..., x_n; c_1,..., c_k, \xi\right) =$$
$$= f_1^{k-1}\Big(t, y_1,..., y_{k-1}, y_k + \varphi_k^{k-1}\left(t, y_1,..., y_{k-1}, x_{k+1},..., x_n; c_1,..., c_k, \xi\right), x_{k+1},..., x_n; c_1,..., c_{k-1,}\xi\Big),...,$$
$$f_{k-1}^k\left(t, y_1,..., y_k, x_{k+1},..., x_n; c_1,..., c_k, \xi\right) =$$
$$= f_{k-1}^{k-1}\Big(t, y_1,..., y_{k-1}, y_k + \varphi_k^{k-1}\left(t, y_1,..., y_{k-1}, x_{k+1},..., x_n; c_1,..., c_k, \xi\right), x_{k+1},..., x_n; c_1,..., c_{k-1,}\xi\Big),$$

$$f_k^k\left(t, y_1,..., y_k, x_{k+1},..., x_n; c_1,..., c_k, \xi\right) =$$
$$= f_k^{k-1}\Big(t, y_1,..., y_{k-1}, y_k + \varphi_k^{k-1}\left(t, y_1,..., y_{k-1}, x_{k+1},..., x_n; c_1,..., c_k, \xi\right), x_{k+1},..., x_n; c_1,..., c_{k-1,}\xi\Big) -$$
$$-\frac{\partial\varphi_k^{k-1}\left(t, y_1,..., y_{k-1}, x_{k+1},..., x_n; c_1,..., c_k, \xi\right)}{\partial t} - \sum_{j=1}^{k-1}\frac{\partial\varphi_k^{k-1}\left(t, y_1,..., y_{k-1}, x_{k+1},..., x_n; c_1,..., c_k, \xi\right)}{\partial y_j}\times$$
$$\times f_j^k\left(t, y_1,..., y_k, x_{k+1},..., x_n; c_1,..., c_k, \xi\right) - \sum_{i=k+1}^{n}\frac{\partial\varphi_k^{k-1}\left(t, y_1,..., y_{k-1}, x_{k+1},..., x_n; c_1,..., c_k, \xi\right)}{\partial x_i}\times \quad ,$$
$$\times f_i^k\left(t, y_1,..., y_k, x_{k+1},..., x_n; c_1,..., c_k, \xi\right),\ f_{k+1}^k\left(t, y_1,..., y_k, x_{k+1},..., x_n; c_1,..., c_k, \xi\right) =$$
$$= f_{k+1}^{k-1}\Big(t, y_1,..., y_{k-1}, y_k + \varphi_k^{k-1}\left(t, y_1,..., y_{k-1}, x_{k+1},..., x_n; c_1,..., c_k, \xi\right), x_{k+1},..., x_n; c_1,..., c_{k-1,}\xi\Big),...,$$
$$f_n^k\left(t, y_1,..., y_k, x_{k+1},..., x_n; c_1,..., c_k, \xi\right) =$$
$$= f_n^{k-1}\Big(t, y_1,..., y_{k-1}, y_k + \varphi_k^{k-1}\left(t, y_1,..., y_{k-1}, x_{k+1},..., x_n; c_1,..., c_k, \xi\right), x_{k+1},..., x_n; c_1,..., c_{k-1,}\xi\Big).$$

The $X_{t_0}^{y_1...y_{k-1}}$ transforms into $X_{t_0}^{y_1...y_k} = \theta_k\left(X_{t_0}^{y_1...y_{k-1}}\right)$ that is the orthogonal projection of $X_{t_0}^{y_1...y_{k-1}}$ onto $R^{2n+m}_{t, y_1,..., y_{k-1}, x_{k+1},..., x_n, c, \xi}$, that is $X_{t_0}^{y_1...y_k} \subseteq R^{2n+m}_{t, y_1,..., y_{k-1}, x_{k+1},..., x_n, c, \xi}$.

Thus at the end of the second stage of the cascade of flattening diffeomorphisms we receive the following whole after-($k$)th-stage-of-cascade-of-diffeomorphisms system

$$\left\{\begin{array}{l}\left\{\begin{array}{l}\Gamma_{x_i}^{k}\left(t,y_1,...,y_k,x_{k+1},...,x_n;c_1,...,c_k,\mathrm{c}_i,\xi\right),\pi_{x_i}^{k}\left(c_1,...,c_3,\xi\right),R^1\times\Xi,\\ \mathrm{G}_{\hat{c},\hat{\xi}}^{x_i}\left(t,y_1,...,y_k,x_{k+1},...,x_n\right)^k\end{array}\right\}_{i=1}^{n};\\ \left\{\begin{array}{l}\left\{\begin{array}{l}\dfrac{dy_1}{dt}\in\bigcup\limits_{\xi\in\Xi}f_1^{k}\left(t,y_1,...,y_k,x_{k+1},...,x_n;c_1,...,c_k,\xi\right),\\ \text{.................................................................}\\ \dfrac{dy_3}{dt}\in\bigcup\limits_{\xi\in\Xi}f_k^{k}\left(t,y_1,...,y_k,x_{k+1},...,x_n;c_1,...,c_k,\xi\right),\end{array}\right.\\ \left.\begin{array}{l}\dfrac{dx_4}{dt}\in\bigcup\limits_{\xi\in\Xi}f_{k+1}^{k}\left(t,y_1,...,y_k,x_{k+1},...,x_n;c_1,...,c_k,\xi\right),\\ \text{.................................................................}\\ \dfrac{dx_n}{dt}\in\bigcup\limits_{\xi\in\Xi}f_n^{k}\left(t,y_1,...,y_k,x_{k+1},...,x_n;c_1,...,c_k,\xi\right)\end{array}\right\}\end{array}\right.\end{array}\right\}, \tag{51}$$

where

$$\begin{array}{l}f_1^{k}\left(t,y_1,...,y_k,x_{k+1},...,x_n;c_1,...,c_k,\xi\right)\Big|_{\forall(t,y_1,...,y_k,x_{k+1},...,x_n;c,\xi)\in\Gamma_{x_1}^{k}\left(t,y_1,...,y_k,x_{k+1},...,x_n;c_1,...,c_k,c_i,\xi\right)=\{y_1=0\}}\equiv 0,\ y_{1,0}=0\\ \text{.....................................................................................................................}\ .\\ f_k^{k}\left(t,y_1,...,y_k,x_{k+1},...,x_n;c_1,...,c_k,\xi\right)\Big|_{\forall(t,y_1,...,y_k,x_{k+1},...,x_n;c,\xi)\in\Gamma_{x_k}^{k}\left(t,y_1,...,y_k,x_{k+1},...,x_n;c_1,...,c_k,c_i,\xi\right)=\{y_k=0\}}\equiv 0,\ y_{k,0}=0\end{array}$$

### ***The ($n$)-th stage of the cascade***

From the whole after-($n-1$)th-stage-of-cascade-of-diffeomorphisms system the flattening diffeomorphisms $\theta_n$ transforms

1) the set of its fiber metabundles into the set of following ones

$$\left\{\Gamma_{x_i}^{n}\left(t,y_1,...,y_n;c_1,...,c_n,,\xi\right),\pi_{x_i}^{n}\left(c_1,...,c_n,\xi\right),R^1\times\Xi,\mathrm{G}_{\hat{c},\hat{\xi}}^{x_i}\left(t,y_1,...,y_n\right)^n\right\}_{i=1}^{n} \tag{52}$$

with the total spaces being

$$\left\{\begin{array}{l}\Gamma_{x_1}^{n}\left(t,y_1,...,y_n;c_1,...,c_n,\xi\right)=\{y_1=0\},\\ \text{.......................................................}\\ \Gamma_{x_n}^{n}\left(t,y_1,...,y_n;c_1,...,c_n,\xi\right)=\{y_n=0\}\end{array}\right\}. \tag{53}$$

Here we assume that

$$\left. \begin{aligned} & g_n^{n-1}\left(t, y_1, ..., y_{n-1}, y_n + \varphi_n^{n-1}\left(t, y_1, ..., y_{n-1}; c_1, ..., c_n, \xi\right); c_1, ..., c_{n-1,} \xi\right) - \\ & -c_n \equiv 0 \end{aligned} \right|_{y_n = 0} ; \tag{54}$$

2) its differential inclusion into the differential inclusion

$$\left\{ \begin{array}{l} \dfrac{dy_1}{dt} \in \bigcup\limits_{\xi \in \Xi} f_1^n\left(t, y_1, ..., y_n; c, \xi\right), \\ \ldots\ldots\ldots\ldots\ldots\ldots\ldots\ldots\ldots\ldots\ldots\ldots \\ \dfrac{dx_n}{dt} \in \bigcup\limits_{\xi \in \Xi} f_n^n\left(t, y_1, ..., y_n; c, \xi\right) \end{array} \right\} \Leftrightarrow \left\{ \frac{dy}{dt} \in \bigcup_{\xi \in \Xi} f^n\left(t, y; c, \xi\right) \right\}, \tag{55}$$

where $f^n\left(t, y; c, \xi\right) = \left( f_1^n\left(t, y; c, \xi\right), ..., f_n^n\left(t, y; c, \xi\right)\right), y = \left(y_1, ..., y_n\right)$,

$$\left\{ \begin{aligned} & f_i^n\left(t, y_1, ..., y_n, ; c, \xi\right) = \\ & = f_i^{n-1}\left(t, y_1, ..., y_{n-1}, y_n + \varphi_n^{n-1}\left(t, y_1, ..., y_{n-1}; c_1, ..., c_n, \xi\right); c_1, ..., c_{n-1,} \xi\right) \end{aligned} \right\}_{i=1}^{n-1},$$

$$f_n^n\left(t, y_1, ..., y_n, ; c, \xi\right) = f_n^{n-1}\left(t, y_1, ..., y_{n-1}, y_n + \varphi_n^{n-1}\left(t, y_1, ..., y_{n-1}; c_1, ..., c_n, \xi\right); c_1, ..., c_{n-1,} \xi\right) -$$

$$-\frac{\partial \varphi_n^{n-1}\left(t, y_1, ..., y_{n-1}; c_1, ..., c_n, \xi\right)}{\partial t} - \sum_{j=1}^{n-1} \frac{\partial \varphi_n^{n-1}\left(t, y_1, ..., y_{n-1}; c_1, ..., c_n, \xi\right)}{\partial y_j} f_j^n\left(t, y_1, ..., y_n, ; c, \xi\right).$$

The $X_{t_0}^{y_1 \cdots y_{n-1}}$ transforms into $X_{t_0}^{y_1 \cdots y_n} = \theta_k\left(X_{t_0}^{y_1 \cdots y_{n-1}}\right)$ that is the orthogonal projection of $X_{t_0}^{y_1 \cdots y_{n-1}}$ onto $R^{2n+m}_{t, y_1, ..., y_{n-1}, c, \xi}$, that is $X_{t_0}^{y_1 \cdots y_n} \subseteq R^{2n+m}_{t, y_1, ..., y_{n-1}, c, \xi}$. Actually, after all the transformations we have obtained $X_{t_0}^{y_1 \cdots y_n} = 0$.

If we fix the value of the vector of parameter $\xi = \hat{\xi}$ then the differential inclusion obtained after the ($n$)th-stage of the cascade of flattening diffeomorphisms

$$\frac{dy}{dt} \in \bigcup_{\xi \in \Xi} f^n\left(t, y; c, \xi\right), \tag{56}$$

which we will call the canonical parametric differential inclusion, turns into the free dynamic system

$$\frac{dy}{dt} = f^n\left(t, y; c, \hat{\xi}\right), \tag{57}$$

which is the restriction of the differential inclusion (56) to $T \times R_y^n$. All its integral curves are identical and equal to the ray $T \subset R_t^1 \subset R_{y,t}^{n+1}$, namely

$$y_t\left(y_0,\hat{\xi}\right)=\left(t,y\left(t;y_0,\hat{\xi}\right)\right)\equiv\left(t,\left(\underbrace{y_1\left(t;y_0,\hat{\xi}\right)}_{\substack{\parallel\\0}},...,\underbrace{y_n\left(t;y_0,\hat{\xi}\right)}_{\substack{\parallel\\0}}\right)\right)=\left(t,\left(0,...,0\right)=\mathbf{0}\right)\forall t\in T. \quad (58)$$

***Remark 4.*** *For the free dynamic system (57) all the phase components of all the initial points of its integral curves are equal to zero according to (68) that is*

$$y_{t_0}=\left(t=t_0,y_0=\left(y_{1,0},...,y_{n,0}\right)=\left(0,...,0\right)=\mathbf{0}\right). \quad (59)$$

*The different values of these initial points are taken into consideration through the constant* $c$ *in the right-hand sides of the differential equations of (57) and the one-to-one correspondence between the initial point of any integral curve in* $T \times R_x^n$ *and the initial point of its counterpart in* $T \times R_y^n$ *is given by means of the relation (22), namely*

$$\begin{pmatrix}\left\{g_i\left(t_0,x_0;\xi\right)=c_i\right\}_{i=1}^n\\ \Updownarrow\\ g\left(t_0,x_0;\xi\right)=c\end{pmatrix}\Rightarrow x_0=\psi\left(t_0,c,\xi\right).$$

Thus at the end of the ($n$)th-stage of the cascade of flattening diffeomorphisms we receive the following whole after-($n$)th-stage-of-cascade-of-diffeomorphisms system

$$\left\{\begin{array}{l}\left\{\begin{array}{l}\Gamma_{x_i}^n\left(t,y_1,...,y_n;c_1,...,c_n,\xi\right),\\ \pi_{x_i}^n\left(c_1,...,c_n,\xi\right),R^1\times\Xi,\\ \mathrm{G}_{\hat{c},\hat{\xi}}^{x_i}\left(t,y_1,...,y_n\right)^n\end{array}\right\}_{i=1}^n;\\ \left\{\begin{array}{l}\dfrac{dy_1}{dt}\in\bigcup\limits_{\xi\in\Xi}f_1^n\left(t,y_1,...,y_n;c,\xi\right),\\ \dots\dots\dots\dots\dots\dots\dots\dots\dots\dots\\ \dfrac{dx_n}{dt}\in\bigcup\limits_{\xi\in\Xi}f_n^n\left(t,y_1,...,y_n;c,\xi\right)\end{array}\right\}\end{array}\right\}\Leftrightarrow\left\{\begin{array}{l}\left\{\Gamma_{x_i}^n\left(t,y;c,\xi\right),\pi_{x_i}^n\left(c,\xi\right),R^1\times\Xi,\mathrm{G}_{\hat{c},\hat{\xi}}^{x_i}\left(t,y\right)^n\right\}_{i=1}^n;\\ \left\{\dfrac{dy}{dt}\in\bigcup\limits_{\xi\in\Xi}f^n\left(t,y;c,\xi\right)\right\}\end{array}\right\}, \quad (60)$$

where $\left\{ f_i^n\left(t,y;c,\xi\right)\Big|_{\forall\left(t,y;c,\xi\right)\in\Gamma^n_{x_i}\left(t,y,c,\xi\right)=\left\{y_i=0\right\}} \equiv 0, y_{i,0}=0\right\}_{i=1}^{n}$.

The cascade of sequential flattening diffeomorphisms $\theta$ creates the diffeomorphism $\overline{\varphi}$ that establishes the diffeomorphic correspondence between the initial system (30) and the ultimately-transformed one (60). Let us find it using the inverse cascade of sequential substitutions as follows

$$\left\{\begin{array}{l}
x_n = y_n + \varphi_n^{n-1}\left(t,y_1,...,y_{n-1};c_1,...,c_n,\xi\right) = \overline{\varphi}_n^{n-1}\left(t,y;c,\xi\right) \\
\downarrow \\
\downarrow -------------\rightarrow \underset{x_n=\overline{\varphi}_n^{n-1}(\cdot)}{\overset{x_n}{\downarrow}} \\
x_{n-1} = y_{n-1} + \varphi_{n-1}^{n-2}\left(t,y_1,...,y_{n-2},x_n;c_1,...,c_{n-1},\xi\right) = \\
\downarrow = y_{n-1} + \varphi_{n-1}^{n-2}\left(t,y_1,...,y_{n-2},\overline{\varphi}_n^{n-1}\left(t,y;c,\xi\right);c_1,...,c_{n-1},\xi\right) = \overline{\varphi}_{n-1}^{n-2}\left(t,y;c,\xi\right) \\
\downarrow \\
\downarrow -------------\rightarrow \underset{x_{n-1}=\overline{\varphi}_{n-1}^{n-2}(\cdot)}{\overset{x_{n-1}}{\downarrow}} \quad \underset{x_n=\overline{\varphi}_n^{n-1}(\cdot)}{\overset{x_n}{\downarrow}} \\
x_{n-2} = y_{n-2} + \varphi_{n-2}^{n-3}\left(t,y_1,...,y_{n-3},x_{n-1},x_n;c_1,...,c_{n-2},\xi\right) = \\
\downarrow = y_{n-2} + \varphi_{n-2}^{n-3}\left(t,y_1,...,y_{n-3},\overline{\varphi}_{n-1}^{n-2}\left(t,y;c,\xi\right),\overline{\varphi}_n^{n-1}\left(t,y;c,\xi\right);c_1,...,c_{n-2},\xi\right) = \overline{\varphi}_{n-2}^{n-3}\left(t,y;c,\xi\right) \\
\downarrow \\
\downarrow ------------\rightarrow \underset{x_n=\overline{\varphi}_{n-2}^{n-3}(\cdot)}{\overset{x_{n-2}}{\downarrow}} -- \underset{x_{n-1}=\overline{\varphi}_{n-1}^{n-2}(\cdot)}{\overset{x_{n-1}}{\downarrow}} -- \underset{x_n=\overline{\varphi}_n^{n-1}(\cdot)}{\overset{x_n}{\downarrow}} \\
\cdots\cdots\cdots\cdots\cdots\cdots\cdots\cdots\cdots\cdots\cdots\cdots\cdots\cdots\cdots\cdots \\
\downarrow \\
\downarrow ------------\rightarrow \underset{x_n=\overline{\varphi}_{k+1}^{k}(\cdot)}{\overset{x_{k+1}}{\downarrow}} -- \underset{x_n=\overline{\varphi}_n^{n-1}(\cdot)}{\overset{x_n}{\downarrow}} \\
x_k = y_k + \varphi_k^{k-1}\left(t,y_1,...,y_{k-1},x_{k+1},.......,x_n;c_1,...,c_k,\xi\right) = \\
\downarrow \\
\downarrow = y_k + \varphi_k^{k-1}\left(t,y_1,...,y_{k-1},\overline{\varphi}_{k+1}^{k}\left(t,y;c,\xi\right),...,\overline{\varphi}_n^{n-1}\left(t,y;c,\xi\right);c_1,...,c_k,\xi\right) = \overline{\varphi}_k^{k-1}\left(t,y;c,\xi\right) \\
\downarrow \\
\downarrow ------------\rightarrow \underset{x_k=\overline{\varphi}_k^{k-1}(\cdot)}{\overset{x_k}{\downarrow}} ---- \underset{x_n=\overline{\varphi}_n^{n-1}(\cdot)}{\overset{x_n}{\downarrow}} \\
\cdots\cdots\cdots\cdots\cdots\cdots\cdots\cdots\cdots\cdots\cdots\cdots\cdots\cdots\cdots\cdots \\
\downarrow \\
\downarrow ----\rightarrow \underset{x_2=\overline{\varphi}_2^{1}(\cdot)}{\overset{x_2}{\downarrow}} -- \underset{x_n=\overline{\varphi}_n^{n-1}(\cdot)}{\overset{x_n}{\downarrow}} \\
x_1 = y_1 + \varphi_1\left(t,x_2,.......,x_n;c_1,\xi\right) = \varphi_1\left(t,\overline{\varphi}_2^{1}\left(t,y;c,\xi\right),...,\overline{\varphi}_n^{n-1}\left(t,y;c,\xi\right);c_1,\xi\right) = \overline{\varphi}_1^{0}\left(t,y;c,\xi\right)
\end{array}\right\}. \quad (61)$$

Thus we receive

$$\bar{\varphi}=\left(\bar{\varphi}_1^0(t,y;c,\xi),\bar{\varphi}_2^1(t,y;c,\xi),\ldots,\bar{\varphi}_k^{k-1}(t,y;c,\xi),\ldots,\bar{\varphi}_{n-1}^{n-2}(t,y;c,\xi),\bar{\varphi}_n^{n-1}(t,y;c,\xi)\right); \quad (62)$$

$$\bar{\varphi}:\left\{\begin{array}{l}\left\{\begin{array}{l}\Gamma_{x_i}^n(t,y;c,\xi),\pi_{x_i}^n(c,\xi),\\ R^1\times\Xi,\mathrm{G}_{\hat{c},\hat{\xi}}^{x_i}(t,y)^n\end{array}\right\}_{i=1}^n;\\ \left\{\dfrac{dy}{dt}\in\bigcup\limits_{\xi\in\Xi}f^n(t,y;c,\xi),\right\}\end{array}\right\}\rightarrow\left\{\begin{array}{l}\left\{\begin{array}{l}\Gamma_{x_i}(t,x;\mathrm{c}_i,\xi),\pi_{x_i}(\mathrm{c}_i,\xi),\\ R^1\times\Xi,\mathrm{G}_{\hat{c}_i,\hat{\xi}}^{x_i}(t,x)\end{array}\right\}_{i=1}^n;\\ \dfrac{dx}{dt}\in\bigcup\limits_{\xi\in\Xi}f(t,x;\xi)\end{array}\right\} \quad (63)$$

or

$$\left\{\begin{array}{l}\left\{\begin{array}{l}\Gamma_{x_i}(t,x;\mathrm{c}_i,\xi),\pi_{x_i}(\mathrm{c}_i,\xi),\\ R^1\times\Xi,\mathrm{G}_{\hat{c}_i,\hat{\xi}}^{x_i}(t,x)\end{array}\right\}_{i=1}^n;\\ \dfrac{dx}{dt}\in\bigcup\limits_{\xi\in\Xi}f(t,x;\xi)\end{array}\right\}\xrightarrow{\theta:\ \theta_1\Rightarrow\ldots\Rightarrow\theta_k\Rightarrow\ldots\Rightarrow\theta_n}\left\{\begin{array}{l}\left\{\begin{array}{l}\Gamma_{x_i}^n(t,y;c,\xi),\pi_{x_i}^n(c,\xi),\\ R^1\times\Xi,\mathrm{G}_{\hat{c},\hat{\xi}}^{x_i}(t,y)^n\end{array}\right\}_{i=1}^n;\\ \left\{\dfrac{dy}{dt}\in\bigcup\limits_{\xi\in\Xi}f^n(t,y;c,\xi)\right\}\end{array}\right\}. \quad (64)$$

**Definition 2.** The system of the relations (60) is called the canonical form of the representation of the initial system (30) and the diffeomorphism $\bar{\varphi}$ establishing the diffeomorphic correspondence between the initial system (30) and its canonical form is called the canonizing diffeomorphism.

Fig. 4 illustrates the action of the cascade of sequential flattening diffeomorphisms $\theta$ and of the canonizing diffeomorphism $\bar{\varphi}$.

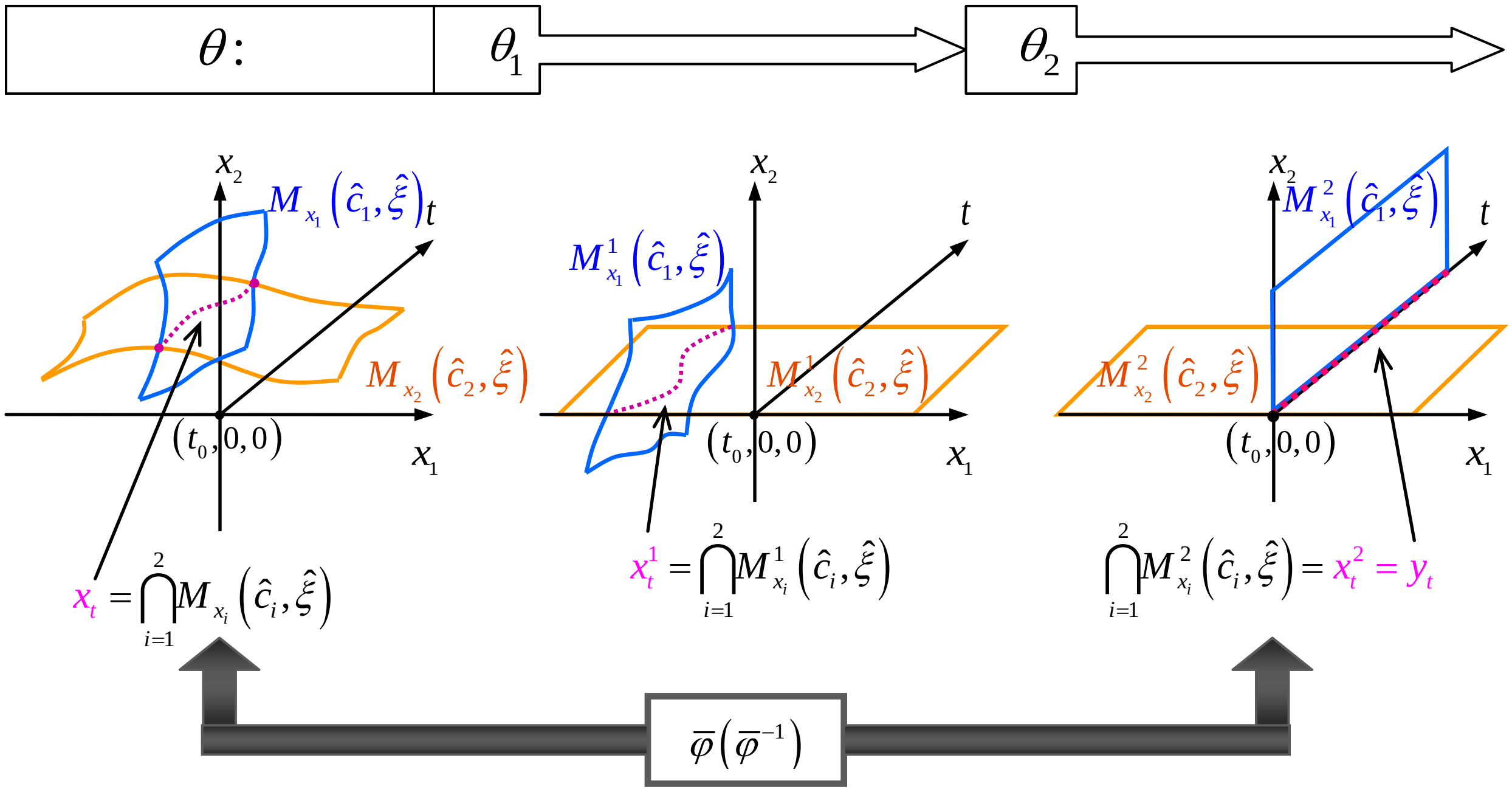

**FIG. 4**

**Definition 3**. We say that a given parametric function $\overline{\varphi}_i^{i-1}\left(t,y;c,\xi\right), i\in\left(1,...,n\right)$ of a real variable $t\in T$ is uniformly convergent in $y$ to $\overline{\varphi}_i^{i-1}\left(t,0;c,\xi\right)$ if for every $\varepsilon>0$ there exists a such neighborhood $Y\left(\varepsilon\right)$ of the point $\hat{y}=0$ that for any $y\in Y\left(\varepsilon\right)$ and $\forall\left(t,c,\xi\right)\in T\times R^n\times\Xi$ we have

$$\left\|\overline{\varphi}_i^{i-1}\left(t,y;c,\xi\right)-\overline{\varphi}_i^{i-1}\left(t,0;c,\xi\right)\right\|<\varepsilon\,, \tag{65}$$

where $y=\left(y_1,...,y_n\right)$ is considered a vector of parameter in this definition.

We will designate the uniform convergence of $\overline{\varphi}_i^{i-1}\left(t,y;c,\xi\right)$ as

$$\overline{\varphi}_i^{i-1}\left(t,y;c,\xi\right)\overbrace{---\rightarrow}^{y\rightarrow 0}\overline{\varphi}_i^{i-1}\left(t,0;c,\xi\right). \tag{66}$$

**Assumption.** All considered spaces and manifolds before and after applying diffeomorphisms intersect transversely.

This means that all spaces and manifolds are in general positions without any special cases of degeneration, topological pathology, tangency, etc.

**Lemma 1.** The geometrical-topological relation between the total spaces of the fiber metabundle (5) $\left\{\Gamma_x\left(t,x;x_0,\xi\right),\pi_x\left(x_0,\xi\right),X_{t_0}\times\Xi,\mathrm{G}_{\hat{x}_0,\hat{\xi}}\left(t,x\right)\right\}$ and the set of the fiber metabundles from (30) $\left\{\Gamma_{x_i}\left(t,x;\mathrm{c}_i,\xi\right),\pi_{x_i}\left(\mathrm{c}_i,\xi\right),R^1\times\Xi,\mathrm{G}_{\hat{c}_i,\hat{\xi}}^{x_i}\left(t,x\right)\right\}_{i=1}^n$ is described with the following diagrams

1. For the total spaces

$$\begin{array}{ccc} & \Gamma_x\left(t,x;x_0,\xi\right)=\bigcap\limits_{i=1}^{n}\Gamma_{x_i}\left(t,x;\mathrm{c}_i,\xi\right) & \\ \nwarrow\!\!\searrow & & \swarrow\!\!\nearrow \\ & \psi\left(\psi^{-1}\right):R^n\times\Xi\rightleftarrows X_{t_0} & \end{array}\;; \tag{67}$$

2. For the typical fibers

$$\begin{array}{ccc} & \Gamma_x\left(t,x;\hat{x}_0,\hat{\xi}\right)=\bigcap\limits_{i=1}^{n}\Gamma_{x_i}\left(t,x;\hat{\mathrm{c}}_i,\hat{\xi}\right)=x_t\left(\hat{x}_0,\hat{\xi}\right) & \\ \nwarrow\!\!\searrow & & \swarrow\!\!\nearrow \\ & \psi\left(\psi^{-1}\right):\left(\hat{\mathrm{c}}_i,\hat{\xi}\right)\rightleftarrows\hat{x}_0 & \end{array}\,. \tag{68}$$

Deciphering analytically the diagram for the typical fibers of the fiber metabundles we have

$$x_t\left(\hat{x}_0,\hat{\xi}\right)=\left\{g\left(t,x;\hat{\xi}\right)=\hat{c};\forall\left(t,x\right)\in T\times R_x^n,\forall\left(\hat{c},\xi\right)\in R^n\times\Xi\right\}, \tag{69}$$

where

$$\hat{x}_0=\psi\left(t_0,\hat{c},\hat{\xi}\right). \tag{70}$$

In other words a specific integral curve $x_t\left(\hat{x}_0,\hat{\xi}\right)$ is the intersection of $n$ of 1-codimensional invariant manifolds $\left\{M_{x_i}\left(\hat{c}_i,\hat{\xi}\right)\right\}_{i=1}^n$ that are the level sections of the first integrals $g\left(t,x;\hat{\xi}\right)=c$ with the hyperplanes $c=\hat{c}$. The correspondence between the initial point $x_{t_0}\left(\hat{x}_0,\hat{\xi}\right)=\left(t_0,\hat{x}_0\right)$ of the integral curve and the numbers $\hat{c}$ is established by the equation (70).

**Proposition.** The canonizing diffeomorphism $\overline{\varphi}$, conserving the intersections and their transversality**,** has the following properties

$$\begin{gathered}\overline{\varphi}\left(\bigcap_{i=1}^n\Gamma_{x_i}^n\left(t,y;c,\xi\right)\right)\equiv\left|\bigcap_{i=1}^n\Gamma_{x_i}^n\left(t,y;c,\xi\right)=\left\{y_i=0\right\}_{i=1}^n\right|=\\ =\bigcap_{i=1}^n\overline{\varphi}\left(\Gamma_{x_i}^n\left(t,y;c,\xi\right)\right)=\bigcap_{i=1}^n\Gamma_{x_i}\left(t,x;\mathrm{c}_i,\xi\right)=\Gamma_x\left(t,x;x_0,\xi\right)\\ \searrow\qquad\qquad\nearrow\\ \psi\left(\psi^{-1}\right):R^n\times\Xi\rightleftarrows X_{t_0}\end{gathered} \tag{71}$$

and

$$\begin{gathered}\overline{\varphi}\left(\bigcap_{i=1}^n\Gamma_{x_i}^n\left(t,y;\hat{c},\hat{\xi}\right)\right)\equiv\left|\bigcap_{i=1}^n\Gamma_{x_i}^n\left(t,y;\hat{c},\hat{\xi}\right)=\left\{y_i=0\right\}_{i=1}^n=y_t\left(0,\hat{\xi}\right)\right|=\\ =\bigcap_{i=1}^n\overline{\varphi}\left(\Gamma_{x_i}^n\left(t,y;\hat{c},\hat{\xi}\right)\right)=\bigcap_{i=1}^n\Gamma_{x_i}\left(t,x;\hat{\mathrm{c}}_i,\hat{\xi}\right)=\Gamma_x\left(t,x;\hat{x}_0,\hat{\xi}\right)=x_t\left(\hat{x}_0,\hat{\xi}\right)\ .\\ \searrow\qquad\qquad\nearrow\\ \psi\left(\psi^{-1}\right):\left(\hat{\mathrm{c}}_i,\hat{\xi}\right)\rightleftarrows\hat{x}_0\end{gathered} \tag{72}$$

Consider the set of the coverings $\left\{\left(p_i\left(c,\xi\right),T\times R_y^n\times R^n\times\Xi,S_i\right)\right\}_{i=1}^n$, consisting of the covering spaces $\left\{S_i\right\}_{i=1}^n$, the base spaces $\left\{T\times R_y^n\times R^n\times\Xi\right\}_{i=1}^n$ and the covering mappings $\left\{\left(p_i(c,\xi):T\times R_y^n\times R^n\times\Xi\to S_i\right)\right\}_{i=1}^n$, which is generated by the equations of the differential inclusion of the total system written in the canonical form (60), namely

$$\left\{p_i(c,\xi):T\times R_y^n\times R^n\times\Xi\to S_i=\left\{s_i=f_i^n(t,y;c,\xi);(t,y)\in T\times R_y^n,(c,\xi)\in R^n\times\Xi\right\}\right\}_{i=1}^n, \quad (73)$$

where $\left\{f_i^n(t,y;c,\xi)\Big|_{\forall(t,y;c,\xi)\in\Gamma_{x_i}^n(t,y,c,\xi)=\{y_i=0\}}\equiv 0, y_{i,0}=0\right\}_{i=1}^n$.

**Definition 4**. The mapping $p(c,\xi)=(p_1(c,\xi),...,p_n(c,\xi))$ is called the total covering mapping of the collection of all the coverings $\left\{\left(p_i(c,\xi),T\times R_y^n\times R^n\times\Xi,S_i\right)\right\}_{i=1}^n$.

**Theorem 1**. Consider the initial system (30) and its canonical representation (60) obtained by means of the canonizing diffeomorphism $\overline{\varphi}$. Let

1. $p(\hat{c},\hat{\xi})$ being the restriction of total covering mapping $p(c,\xi)$ to $T\times R_y^n$ belongs to the class of $\mathbf{A}$ – mappings, that is $p(\hat{c},\hat{\xi})\in[p]^{\mathbf{A}}$ [4, p. 19];
2. $\overline{\varphi}(t,y;\hat{c},\hat{\xi})$ being the restriction of the canonizing diffeomorphism $\overline{\varphi}(t,y;c,\xi)$ to $T\times R_y^n$ is uniformly convergent to $\overline{\varphi}(t,0;\hat{c},\hat{\xi})$ in $y$,

that is

$$\overline{\varphi}(t,y;\hat{c},\hat{\xi})\overbrace{---\to}^{y\to0}\overline{\varphi}(t,0;\hat{c},\hat{\xi}). \quad (74)$$

Then the canonical form of the positive definite auxiliary functions $W(y)$ of Lyapunov functions $V(t,y)$ for the canonical parametric differential inclusion (60)

$$W(y)=\sum_{i=1}^n y_i^2 \quad (75)$$

ensures

a) the asymptotical stability of the integral curve $y_t(\hat{y}_0,\hat{\xi})=\left(t,y(t;\hat{y}_0,\hat{\xi})\right)\equiv(t,\mathbf{0})\forall t\in T$ starting from the initial point $(t=t_0,\hat{y}_0=\mathbf{0})$ of the free dynamic system

$$\frac{dy}{dt}=f^n(t,y;\hat{c},\hat{\xi}), \quad (76)$$

which is the restriction of the canonical differential inclusion

$$\frac{dy}{dt} \in \bigcup_{\xi \in \Xi} f^{n}\left(t, y; c, \xi\right)$$

to $T \times R_y^n$ of the system (58) and according to the canonizing diffeomorphism $\bar{\varphi}$ ;

b) the asymptotical stability of the integral curves $x_t\left(\hat{x}_0, \hat{\xi}\right)$ starting from the initial point $\hat{x}_0 \in X_{t_0}$ of the free dynamical system (2), namely

$$\frac{dx}{dt} = f\left(t, x; \hat{\xi}\right),$$

which is the restriction of the initial differential inclusion (1) to $T \times R_y^n$, where $\hat{x}_0 = \psi\left(t_0, \hat{c}, \hat{\xi}\right)$ (see (22)).

The Theorem 1 is formulated according to Lyapunov's second method for stability as it is expounded in [5]. The proof of this theorem and all the ones later on in the paper can be easily conducted on the basis of the methodology and techniques developed in [6].

**Corollary 1.1** The $n$-dimensional invariant manifolds $\left\{\Gamma_{x_i}\left(t, x; \hat{c}_i, \hat{\xi}\right)\right\}_{i=1}^{n}$ and the integral curve

$$x_t\left(\hat{x}_0, \hat{\xi}\right) = \Gamma_x\left(t, x; \hat{x}_0, \hat{\xi}\right) = \bigcap_{i=1}^{n} \Gamma_{x_i}\left(t, x; \hat{c}_i, \hat{\xi}\right)$$

are at least mosaic ω-attractors, that is the ones belong to the class of ω-attractors

$$x_t\left(\hat{x}_0, \hat{\xi}\right) = \Gamma_x\left(t, x; \hat{x}_0, \hat{\xi}\right) \in [\mathrm{A}^{\omega}], \Gamma_{x_i}\left(t, x; \hat{c}_i, \hat{\xi}\right) \in [\mathrm{A}^{\omega}].$$

It is not difficult to construct the Lyapunov function on the basis of the auxiliary function $W\left(y\right)$ shown in (75). For example, it can be written as given below

$$V\left(t, y\right) = W\left(y\right)\left(\lambda + e^{-t}\right), \tag{77}$$

where $\lambda \geq 1$.

It is obvious that there exists an infinite number of the functions, which satisfy the conditions of the basic theorem, namely

$$W\left(y\right)=\sum_{i=1}^{n}a_iy_i^l \tag{78}$$

where $a_i \in R^+, l = 2\tilde{l}, \tilde{l} \in N^+$. From the expressions (75) and (78) it is clear why we call the former form of the function $W\left(y\right)$ canonical.

## 3. CLASSIFICATIONAL STABILITY OF TYPICAL FIBER OF FIBER SUBBUNDLES AND METABUNDLES

In this section we will consider the *classificational stability* of the typical fiber, which is a ω-attractor in the structure of fiber metabundle (25) through its two fiber subbundles (26) and (27). The concept of this stability has been introduced in [6, p. 96]. It will open us the possibilities to investigate the classificational stability of the total typical fiber, which is an asymptotically stable integral curve belonging to the class of ω-attractor, of the metabundle (5) through the fiber subbundles (6) and (7). In other words, we would like to see under what conditions the integral curve preserves its asymptotical stability when we play with the vectors of parameters $\xi$ and $c$ or $x_0$.

The typical fiber of the fiber metabundle also as of the fiber subbundles (26) and (27) is a $n$-dimensional invariant manifold $\Gamma_{x_i}\left(t,x;\hat{x}_0,\hat{\xi}\right)\xleftrightarrow{\hat{c}_i=g_i\left(t_0,\hat{x}_0;\hat{\xi}\right)}M_{x_i}\left(\hat{c}_i,\hat{\xi}\right)$.

**I.** The topological and analytical criteria of the classificational stability of the typical fiber of the fiber subbundle (26), namely

$$\left\{\Gamma_{x_i}\left(t,x;\hat{x}_0,\xi\right),\pi_{x_i}\left(\hat{x}_0,\xi\right)\xleftrightarrow{\hat{c}_i=g_i\left(t_0,\hat{x}_0;\xi\right)}\pi_{x_i}\left(\hat{c}_i,\xi\right),\Xi,\mathrm{G}_{\hat{x}_0,\hat{\xi}}^{x_i}\left(t,x\right)\right\}$$

can be formulated on the base of Theorems 3 and 8 of [4]. Further in the paper we will consider the collection of the fiber bundles

$$\left\{\Gamma_{x_i}\left(t,x;\hat{x}_0,\xi\right),\pi_{x_i}\left(\hat{x}_0,\xi\right)\xleftrightarrow{\hat{c}_i=g_i\left(t_0,\hat{x}_0;\xi\right)}\pi_{x_i}\left(\hat{c}_i,\xi\right),\Xi,\mathrm{G}_{\hat{x}_0,\hat{\xi}}^{x_i}\left(t,x\right)\right\}_{i=1}^{n}$$

and the collections of their total subspaces $\left\{\Gamma_{x_i}\left(t,x;\hat{x}_0,\xi\right)\right\}_{i=1}^{n}$ and typical fibers

$$\left\{\Gamma_{x_i}\left(t,x;\hat{x}_0,\hat{\xi}\right)\xleftrightarrow{\hat{c}_i=g_i\left(t_0,\hat{x}_0;\hat{\xi}\right)}M_{x_i}\left(\hat{c}_i,\hat{\xi}\right)\right\}_{i=1}^{n}.$$

**Theorem 2**. Let there exists a $m$-dimensional neighborhood, $\Xi_{\hat{\xi}} \subset \Xi$, of the point $\hat{\xi}$ such that the following condition(s) hold true for all $\xi \in \Xi_{\hat{\xi}}$:

1. Topological criterion:

$$p_i(\hat{c}, \xi) \in [p_i]^{\mathrm{A}}. \tag{79}$$

2. Analytical criterion:

   2.1. The classificational rank $\bar{s}$ for the covering mapping $p_i(\hat{c}, \xi)$ preserves the odd parity.

   2.2. The negative sign of the partial derivative $\left.\left(\dfrac{\partial^{\bar{s}} f_i^n(t, y; \hat{c}, \xi)}{\partial y_i^{\bar{s}}}\right)\right|_{y_i \equiv 0}$ is preserved, that is

$$\left.\boldsymbol{sign}\left(\frac{\partial^{\bar{s}} f_i^n(t, y; \hat{c}, \xi)}{\partial y_i^{\bar{s}}}\right)\right|_{y_i \equiv 0} = \left.\boldsymbol{sign}\left(\frac{\partial^{\bar{s}} f_i^n(t, y; \hat{c}, \hat{\xi})}{\partial y_i^{\bar{s}}}\right)\right|_{y_i \equiv 0}. \tag{80}$$

Then $\left\{\Gamma_{x_i}\left(t, x; \hat{x}_0, \hat{\xi}\right) \xleftarrow{\hat{c}_i = g_i\left(t_0, \hat{x}_0; \hat{\xi}\right)} M_{x_i}\left(\hat{c}_i, \hat{\xi}\right)\right\} \in [\mathrm{A}^{\omega}]$ is classificationally stable at the point $\hat{\xi}$ meaning that

$$\left\{\Gamma_{x_i}\left(t, x; \hat{x}_0, \xi\right) \xleftarrow{\hat{c}_i = g_i\left(t_0, \hat{x}_0; \xi\right)} M_{x_i}\left(\hat{c}_i, \xi\right)\right\} \in [\mathrm{A}^{\omega}] \forall \xi \in \Xi_{\hat{\xi}} \tag{81}$$

**Corollary 2.1** If at the point $\xi = \hat{\xi} \in \Xi$ the classificational rank $\bar{s}$ equals the unit and the sign of the partial derivative $\left.\left(\dfrac{\partial^{\bar{s}} f_i^n(t, y; \hat{c}, \xi)}{\partial y_i^{\bar{s}}}\right)\right|_{y_i \equiv 0}$ is negative, that is

$$\left.\boldsymbol{sign}\left(\frac{\partial^{\bar{s}} f_i^n(t, y; \hat{c}, \hat{\xi})}{\partial y_i^{\bar{s}}}\right)\right|_{y_i \equiv 0} = -1, \bar{s} = 1, \tag{82}$$

then $\left\{\Gamma_{x_i}\left(t, x; \hat{x}_0, \hat{\xi}\right) \xleftarrow{\hat{c}_i = g_i\left(t_0, \hat{x}_0; \hat{\xi}\right)} M_{x_i}\left(\hat{c}_i, \hat{\xi}\right)\right\} \in [\mathrm{A}^{\omega}]$ is always classificationally stable and there is no necessity to satisfy the two above-mentioned preservation laws (2.1) and (2.2) for all points $\xi \in \Xi_{\hat{\xi}}$.

It is known that

$$\Gamma_{\hat{x}_0}\left(t,x;\hat{x}_0,\hat{\xi}\right)=\left\{x_t\left(\hat{x}_0,\hat{\xi}\right)\in X_t\left(\hat{\xi}\right)\right\}=\bigcap_{i=1}^{n}\Gamma_{x_i}\left(t,x;\hat{x}_0,\hat{\xi}\right), \tag{83}$$

where $\Gamma_{\hat{x}_0}\left(t,x;\hat{x}_0,\hat{\xi}\right)$ is the typical fiber being an integral curve of the fiber subbundle (6).

**Corollary 2.2** Let all $n$ typical fibers $\left\{\Gamma_{x_i}\left(t,x;\hat{x}_0,\hat{\xi}\right)\xleftrightarrow{\hat{c}_i=g_i\left(t_0,\hat{x}_0;\hat{\xi}\right)}M_{x_i}\left(\hat{c}_i,\hat{\xi}\right)\right\}\in[\mathrm{A}^{\omega}]$ of the collection $\left\{\Gamma_{x_i}\left(t,x;\hat{x}_0,\hat{\xi}\right)\xleftrightarrow{\hat{c}_i=g_i\left(t_0,\hat{x}_0;\hat{\xi}\right)}M_{x_i}\left(\hat{c}_i,\hat{\xi}\right)\right\}_{i=1}^{n}$ are classificationally stable at the point $\hat{\xi}\in\Xi_{\hat{\xi}}$, then the typical fiber (83) being the integral curve $x_t\left(\hat{x}_0,\hat{\xi}\right)$ asymptotically stable at the point $\hat{\xi}$ is also classificationally stable at this point.

In geometrical terms, all integral curves $x_t\left(\hat{x}_0,\xi\right)$ composing the bouquet $\bigcup_{\xi\in\Xi_{\hat{\xi}}}x_t\left(\hat{x}_0,\xi\right)$, each of which corresponds to its own point $\xi\in\Xi_{\hat{\xi}}$ and comes out from the common point $x_{t_0}=\left(t_0,\hat{x}_0\right)$, are asymptotically stable. In topological terms, the property of asymptotical stability is preserved for the restriction of the total subspace of the fiber subbundle (6)

$\Gamma_x\left(t,x;\hat{x}_0,\xi\right)=\left\{x_t\left(\hat{x}_0,\xi\right)=\left(t,x\left(t;\hat{x}_0,\xi\right)\right)\forall\xi\in\Xi\Leftrightarrow\bigcup_{\forall\hat{\xi}\in\Xi}x_t\left(\hat{x}_0,\hat{\xi}\right),\left(t,x\right)\in T\times R_x^n;\hat{x}_0\in X_{t_0}\right\}$ to $\Xi_{\hat{\xi}}\subset\Xi$.

**Definition 5**. If the conditions of Corollary 2.2 holds true for $\forall\xi\in\Xi$ then the asymptotically stable integral curve $x_t\left(\hat{x}_0,\hat{\xi}\right)$ is called classificationally stable in the entire parameter manifold $\Xi$.

**Definition 6**. If the asymptotically stable integral curves $x_t\left(x_0,\hat{\xi}\right)$ are classificationally stable in the entire parameter manifold $\Xi$ and for $\forall x_0\in X_{t_0}\xleftrightarrow{c=g\left(t_0,x_0;\xi\right)\Leftrightarrow x_0=\psi\left(t_0,c,\xi\right)}\forall c\in R^n$ then the parametric differential inclusion (1) is called the one having the global asymptotical stability.

**II.** Now consider the classificational stability of the typical fiber

$\Gamma_{x_i}\left(t,x;\hat{x}_0,\hat{\xi}\right)\xleftrightarrow{\hat{c}_i=g_i\left(t_0,\hat{x}_0;\hat{\xi}\right)}M_{x_i}\left(\hat{c}_i,\hat{\xi}\right)=$

$=\left\{g_i\left(t,x;\hat{\xi}\right)=\hat{c}_i\big|_{\hat{c}_i=g_i\left(t_0,\hat{x}_0;\hat{\xi}\right)} \Leftrightarrow x_i=\varphi_i\left(t,x^i;t_0,\hat{x}_0,\hat{\xi}\right),(t,x)\in T\times R_x^n;\hat{c}_i\in R^1,\hat{\xi}\in\Xi\right\}$ in the structure of the fiber subbundle (27) $\left\{\Gamma_{x_i}\left(t,x;x_0,\hat{\xi}\right),\pi_{x_i}\left(c_i,\hat{\xi}\right),R^1,G^{x_i}_{\hat{x}_0,\hat{\xi}}(t,x)\right\}$ with the total subspace $\Gamma_{x_i}\left(t,x;x_0,\hat{\xi}\right)=F_{x_i}\left(\hat{\xi}\right)=$

$=\left\{\bigcup_{\forall c_i\in R^1}\left\{g_i\left(t,x;\hat{\xi}\right)=c_i\big|_{c_i=g_i\left(t_0,x_0;\hat{\xi}\right)}\Leftrightarrow x_i=\varphi_i\left(t,x^i;t_0,x_0,\hat{\xi}\right)\right\},(t,x)\in T\times R_x^n;\hat{\xi}\in\Xi\right\}$ and the base subspace $R^1$. In order to formulate the conditions of classificational stability in this case we need to turn our attention to the quotient space $X_{t_0}/L_{x_i}\left(\hat{c}_i,\hat{\xi}\right)$, which elements are $(n-1)$-dimensional submanifolds

$$L_{x_i}\left(\hat{c}_i,\hat{\xi}\right)=\left\{\hat{c}_i=g_i\left(t_0,x_0;\hat{\xi}\right)\Rightarrow x_{i,0}=\tilde{\varphi}_i\left(t_0,x_0^i,\hat{\xi};\hat{c}_i\right),x_0\in X_{t_0};\hat{\xi}\in\Xi\right\}$$

of the $n$ - dimensional invariant manifold $\Gamma_{x_i}\left(t,x;\hat{x}_0,\hat{\xi}\right)\xleftrightarrow{\hat{c}_i=g_i\left(t_0,\hat{x}_0;\hat{\xi}\right)}M_{x_i}\left(\hat{c}_i,\hat{\xi}\right)$. The former ones serve the latter ones as the initial points $x_{t_0}$ do the integral curves $x_t$ , that is they are the initial elements from where $\Gamma_{x_i}\left(t,x;\hat{x}_0,\hat{\xi}\right)\xleftrightarrow{\hat{c}_i=g_i\left(t_0,\hat{x}_0;\hat{\xi}\right)}M_{x_i}\left(\hat{c}_i,\hat{\xi}\right)$ start. The quotient space $X_{t_0}/L_{x_i}\left(\hat{c}_i,\hat{\xi}\right)$ can be considered a $(n-1)$-dimensional foliation with $L_{x_i}\left(c_i,\hat{\xi}\right)$ as leaves, where $c_i\in R^1$.

Let us introduce in $X_{t_0}/L_{x_i}\left(\hat{c}_i,\hat{\xi}\right)$ the distance between two arbitrary leaves $L_{x_i}\left(\hat{c}_i'',\hat{\xi}\right)$ and $L_{x_i}\left(\hat{c}_i',\hat{\xi}\right)$ as follows

$$\left\|L_{x_i}\left(\hat{c}_i'',\hat{\xi}\right),L_{x_i}\left(\hat{c}_i',\hat{\xi}\right)\right\|=\inf_{\forall x_0^i\in pr_\perp\left(X_{t_0}\to R_{x^i}^{n-1}\right)}\left|\tilde{\varphi}_i\left(t_0,x_0^i,\hat{\xi};\hat{c}_i''\right)-\tilde{\varphi}_i\left(t_0,x_0^i,\hat{\xi};\hat{c}_i'\right)\right| , \tag{84}$$

where $pr_\perp\left(X_{t_0}\to R_{x^i}^{n-1}\right)$ is the orthogonal projection of the set $X_{t_0}$ to the space $R_{x^i}^{n-1}$. It is obvious that this distance can be also defined in the following way

$$\left\|L_{x_i}\left(\hat{c}_i'',\hat{\xi}\right),L_{x_i}\left(\hat{c}_i',\hat{\xi}\right)\right\|=\left|\hat{c}_i''-\hat{c}_i'\right| . \tag{85}$$

They are two equivalent definitions of the distance between two arbitrary leaves and we can use either of them.

**Definition 7**. The typical fiber $\Gamma_{x_i}\left(t,x;\hat{x}_0,\hat{\xi}\right)\xleftrightarrow{\hat{c}_i=g_i\left(t_0,\hat{x}_0;\hat{\xi}\right)\Rightarrow\hat{x}_{i,0}=\tilde{\varphi}_i\left(t_0,\hat{x}_0^i,\hat{\xi};\hat{c}_i\right)} M_{x_i}\left(\hat{c}_i,\hat{\xi}\right)\in[\mathrm{A}^{\omega}]$ is called classificationally stable in the structure of the fiber subbundle (27) $\left\{\Gamma_{x_i}\left(t,x;x_0,\hat{\xi}\right),\pi_{x_i}\left(c_i,\hat{\xi}\right),R^1,\right.$ $\left.\mathrm{G}_{\hat{x}_0,\hat{\xi}}^{x_i}\left(t,x\right)\right\}$ if there exists a 1-dimensional neighborhood, $C_{\hat{c}_i}\subset R^1$, of the point $\hat{c}$ such that for all $c_i\in C_{\hat{c}_i}$ the following condition holds true

$$\Gamma_{x_i}\left(t,x;x_0,\hat{\xi}\right)\xleftrightarrow{c_i=g_i\left(t_0,x_0;\hat{\xi}\right)\Rightarrow x_{i,0}=\tilde{\varphi}_i\left(t_0,x_0^i,\hat{\xi};c_i\right)} M_{x_i}\left(c_i,\hat{\xi}\right)\in[\mathrm{A}^{\omega}]. \qquad (86)$$

This definition is based on how we have defined the distance between two arbitrary leaves in (85). If we would like to use the relation (84) then we have to replace the final part of the Definition 7 with "…if there exists a $n$-dimensional neighborhood, $EL_{x_i}\left(\hat{c}_i,\hat{\xi}\right)\subset X_{t_0}$, of the a $(n-1)$-dimensional leaf $L_{x_i}\left(\hat{c}_i,\hat{\xi}\right)$ such that for all $L_{x_i}\left(c_i,\hat{\xi}\right)\in EL_{x_i}\left(\hat{c}_i,\hat{\xi}\right)$ the following condition holds true…" If the condition of the classificational stability (86) is valid for all $c_i\in R^1$ or $EL_{x_i}\left(\hat{c}_i,\hat{\xi}\right)\equiv X_{t_0}$, then the total subspace $\Gamma_{x_i}\left(t,x;x_0,\hat{\xi}\right)$ has global classificational stability in the structure of the fiber subbundle (27).

According to Theorem 4 from [4] we have

**Theorem 3**. The typical fiber

$$\Gamma_{x_i}\left(t,x;\hat{x}_0,\hat{\xi}\right)\xleftrightarrow{\hat{c}_i=g_i\left(t_0,\hat{x}_0;\hat{\xi}\right)} M_{x_i}\left(\hat{c}_i,\hat{\xi}\right)=\left\{\begin{array}{l} g_i\left(t,x;\hat{\xi}\right)=\hat{c}_i\big|_{\hat{c}_i=g_i\left(t_0,\hat{x}_0;\hat{\xi}\right)}\Leftrightarrow \\ x_i=\varphi_i\left(t,x^i;t_0,\hat{x}_0,\hat{\xi}\right), \\ \left(t,x\right)\in T\times R_x^n;\hat{c}_i\in R^1,\hat{\xi}\in\Xi \end{array}\right\}\in[\mathrm{A}^{\omega}]$$ of the fiber

subbundle (27) $\left\{\Gamma_{x_i}\left(t,x;x_0,\hat{\xi}\right),\pi_{x_i}\left(c_i,\hat{\xi}\right),R^1,\mathrm{G}_{\hat{x}_0,\hat{\xi}}^{x_i}\left(t,x\right)\right\}$ is always classificationally stable at the point $\hat{x}_0\in X_{t_0}\xleftrightarrow{\hat{c}_i=g_i\left(t_0,\hat{x}_0;\hat{\xi}\right)\Leftrightarrow x_{i,0}=\tilde{\varphi}_i\left(t_0,x_0^i,\hat{\xi};\hat{c}_i\right)}\hat{c}_i\in R^1$ in contrast to ω-repellers and ω-shunts.

**Theorem 4**. Let

$$\Gamma_{\hat{x}_0}\left(t,x;\hat{x}_0,\hat{\xi}\right)=\left\{x_t\left(\hat{x}_0,\hat{\xi}\right)\in X_t\left(\hat{\xi}\right)\right\}=\bigcap_{i=1}^{n}\Gamma_{x_i}\left(t,x;\hat{x}_0,\hat{\xi}\right)$$

and $\left\{\Gamma_{x_i}\left(t,x;\hat{x}_0,\hat{\xi}\right)\right\}_{i=1}^{n}\in[\mathrm{A}^{\omega}]$, then the integral curve $x_t\left(\hat{x}_0,\hat{\xi}\right)$ is asymptotically stable.

There doesn't make sense to talk about the classificational stability of the integral curve $x_t\left(\hat{x}_0,\hat{\xi}\right)$ in this case because it automatically means its asymptotical stability.

**Definition 8**. If all the integral curves of the free dynamic system (2)

$$\frac{dx}{dt}=f\left(t,x;\hat{\xi}\right),$$

that is $\forall x_0=\hat{x}_0\in X_{t_0}$, are asymptotically stable, then the entire system (2) is called asymptotically stable.

**Corollary 4.1**. If all the total subspaces $\left\{\Gamma_{x_i}\left(t,x;x_0,\hat{\xi}\right)\right\}_{i=1}^{n}\in[\mathrm{A}^{\omega}]$ have global classificational stability in the structure of the corresponding fiber subbundles $\left\{\Gamma_{x_i}\left(t,x;x_0,\hat{\xi}\right),\pi_{x_i}\left(c_i,\hat{\xi}\right),R^1,\mathrm{G}_{\hat{x}_0,\hat{\xi}}^{x_i}\left(t,x\right)\right\}_{i=1}^{n}$, then free dynamic system (2)

$$\frac{dx}{dt}=f\left(t,x;\hat{\xi}\right)$$

is asymptotically stable.

**Definition 9**. If in the structure of the parametric differential inclusion (1)

$$\frac{dx}{dt}\in\bigcup_{\xi\in\Xi}f\left(t,x;\xi\right)$$

all the free dynamic system (2)

$$\frac{dx}{dt}=f\left(t,x;\xi\right),$$

that is $\forall\xi\in\Xi$, are asymptotical stable, then the entire inclusion (1) is asymptotical stable.

**Corollary 4.2**. If in the structure of the collection of the fiber metabundle (9)

$$\left\{\left\{\Gamma_x\left(t,x;x_0,\xi\right),\pi_x\left(\hat{x}_0,\xi\right),\Xi,\mathrm{G}_{x_0,\hat{\xi}}\left(t,x\right)\right\}\right\}_{i=1}^{n}$$

all their corresponding typical fibers $\left\{\Gamma_{x_i}\left(t,x;x_0,\hat{\xi}\right)\right\}_{i=1}^{n}\in[\mathrm{A}^{\omega}]$ have global classificational stability and this property is preserved by the section projection 2 $\pi_x\left(\hat{x}_0,\xi\right)$, then the entire parametric differential inclusion (1) is asymptotical stable.

Taking account of

$$\overline{\varphi}:\left\{\begin{array}{l}\left\{\begin{array}{l}\Gamma^n_{x_i}(t,y;c,\xi),\pi^n_{x_i}(c,\xi),\\ R^1\times\Xi,\mathrm{G}^{x_i}_{\hat{c},\hat{\xi}}(t,y)^n\end{array}\right\}^n_{i=1};\\ \left\{\dfrac{dy}{dt}\in\bigcup_{\xi\in\Xi}f^n(t,y;c,\xi),\right\}\end{array}\right\}\rightarrow\left\{\begin{array}{l}\left\{\begin{array}{l}\Gamma_{x_i}(t,x;\mathrm{c}_i,\xi),\pi_{x_i}(\mathrm{c}_i,\xi),\\ R^1\times\Xi,\mathrm{G}^{x_i}_{\hat{c}_i,\hat{\xi}}(t,x)\end{array}\right\}^n_{i=1};\\ \dfrac{dx}{dt}\in\bigcup_{\xi\in\Xi}f(t,x;\xi)\end{array}\right\},$$

Theorem 1. can be expanded over the entire parametric differential inclusion (1).

**Theorem 5**. Let

1. The total covering mapping $p(c,\xi)=\left(p_1(c,\xi),...,p_n(c,\xi)\right)\in[p]^{\mathbf{A}}$;

2. $\overline{\varphi}(t,y;c,\xi)\overbrace{---\rightarrow}^{y\rightarrow 0}\overline{\varphi}(t,0;c,\xi)\forall(c,\xi)\in\mathrm{R}^n\times\Xi.$

Then the canonical form of the positive definite auxiliary functions $W(y)$ of Lyapunov functions $V(t,y)$ for the canonical parametric differential inclusion (56)

$$W(y)=\sum_{i=1}^{n}y_i^2$$

ensures

a) the asymptotical stability of the very (56)

$$\frac{dy}{dt}\in\bigcup_{\xi\in\Xi}f^n(t,y;c,\xi);$$

b) the asymptotical stability of the initial differential inclusion (1)

$$\frac{dx}{dt}\in\bigcup_{\xi\in\Xi}f(t,x;\xi).$$

**Corollary 5.1**. If under the conditions of Theorem 5 we will consider the restriction $p\left(c,\hat{\xi}\right)$ of $p(c,\xi)$ to $T\times R^n_y\times R^n$, then the canonical form of the positive definite auxiliary functions $W(y)$ of Lyapunov functions $V(t,y)$ for the canonical parametric differential inclusion (56)

$$W(y)=\sum_{i=1}^{n} y_i^2$$

insures the asymptotical stability as the free dynamic system (57)

$$\frac{dy}{dt}=f^n\left(t,y;c,\hat{\xi}\right)$$

so the free dynamic system (2)

$$\frac{dx}{dt}=f\left(t,x;\hat{\xi}\right).$$

**Example**

Consider the following differential inclusion

$$\left\{\begin{array}{l} \dfrac{dx_1}{dt}=\dfrac{-\dfrac{d\eta(\bullet)}{dt}\left[2x_1x_2-x_2^2\right]+\dfrac{d\eta(\bullet)}{dt}\gamma(\bullet)+\dfrac{d\gamma(\bullet)}{dt}\eta(\bullet)}{2\eta(\bullet)x_2} \\ \dfrac{dx_2}{dt}=\dfrac{\dfrac{d\gamma(\bullet)}{dt}\eta(\bullet)-\dfrac{d\eta(\bullet)}{dt}\left[x_2^2-\gamma(\bullet)\right]}{2\eta(\bullet)x_2} \end{array}\right\}, \tag{87}$$

where $\eta(\bullet)=\eta(t,\xi_1)$, $\gamma(\bullet)=\gamma(t,\xi_2,\xi_3,\xi_4)$, $\xi=(\xi_1,\xi_2,\xi_3,\xi_4)\in R^4$, $2\eta(\bullet)x_2>0\forall t\in[0,\infty[$.

The complete set of independent parametric first integrals is

$$\left\{\begin{array}{l} (x_1-x_2)\eta(t,\xi_1)=c_1, \\ \dfrac{x_2^2-\gamma(t,\xi_2,\xi_3,\xi_4)}{x_1-x_2}=c_2 \end{array}\right\}. \tag{88}$$

The general solution of the initial differential inclusion (87) is given below

$$\left\{\begin{array}{l} x_1=\dfrac{c_1}{\eta(t,\xi_1)}+\sqrt{\dfrac{c_1c_2}{\eta(t,\xi_1)}+\gamma(t,\xi_2,\xi_3,\xi_4)}, \\ x_2=\sqrt{\dfrac{c_1c_2}{\eta(t,\xi_1)}+\gamma(t,\xi_2,\xi_3,\xi_4)} \end{array}\right\}. \tag{89}$$

In order to avoid making the analysis of the example very complicated, we will not carry out the detailed study of the existence, uniqueness and continuity of the general and particular solutions of the parametric inclusion (87) and respectively we will skip imposing strict conditions on variables and parameters. Our aim is to demonstrate the feasibility and the effectiveness of the technique proposed in the paper.

Thus, the whole initial system is

$$I_x = \left\{ \begin{array}{l} \left\{ \begin{array}{l} \Gamma_{x_1}\left(t,x_1,x_2;c_1,\xi\right) = \left\{ g_1\left(t,x_1,x_2;\xi\right) = \left(x_1 - x_2\right)\eta\left(t,\xi_1\right) = c_1 \right\}, \\ \Gamma_{x_2}\left(t,x_1,x_2;c_2,\xi\right) = \left\{ g_2\left(t,x_1,x_2;\xi\right) = \dfrac{x_2^2 - \gamma\left(t,\xi_2,\xi_3,\xi_4\right)}{x_1 - x_2} = c_2 \right\} \end{array} \right\}; \\ \\ \left\{ \begin{array}{l} \dfrac{dx_1}{dt} = \dfrac{-\dfrac{d\eta(\bullet)}{dt}\left[2x_1x_2 - x_2^2\right] + \dfrac{d\eta(\bullet)}{dt}\gamma(\bullet) + \dfrac{d\gamma(\bullet)}{dt}\eta(\bullet)}{2\eta(\bullet)x_2}, \\ \dfrac{dx_2}{dt} = \dfrac{\dfrac{d\gamma(\bullet)}{dt}\eta(\bullet) - \dfrac{d\eta(\bullet)}{dt}\left[x_2^2 - x_2^2\gamma(\bullet)\right]}{2\eta(\bullet)x_2} \end{array} \right. \end{array} \right\} \tag{90}$$

The cascade of flattening diffeomorphisms consists only of two stages:

$$\theta = \left\{\theta_1 \Rightarrow \theta_2\right\} = \left( \begin{array}{c} x_1 = y_1 + x_2 + \dfrac{c_1}{\eta\left(t,\xi_1\right)} \\ \downarrow \\ x_2 = y_2 + \sqrt{\gamma\left(t,\xi_2,\xi_3,\xi_4\right) + c_2\left[y_1 + \dfrac{c_1}{\eta\left(t,\xi_1\right)}\right]} \end{array} \right). \tag{91}$$

The canonizing diffeomorphism

$$\overline{\varphi}^{-1} = \left\{ \begin{array}{l} x_1 = y_1 + y_2 + \dfrac{c_1}{\eta\left(t,\xi_1\right)} + \sqrt{\gamma\left(t,\xi_2,\xi_3,\xi_4\right) + c_2\left[y_1 + \dfrac{c_1}{\eta\left(t,\xi_1\right)}\right]} \\ x_2 = y_2 + \sqrt{\gamma\left(t,\xi_2,\xi_3,\xi_4\right) + c_2\left[y_1 + \dfrac{c_1}{\eta\left(t,\xi_1\right)}\right]} \end{array} \right\} \tag{92}$$

transforms the initial differential inclusion (87) into

$$\left\{\begin{array}{l}\frac{dy_1}{dt}=-\frac{1}{\eta(\bullet)}\frac{d\eta(\bullet)}{dt}y_1=f_1^2\left(t,y_1,y_2;c_1,c_2,\xi\right),\\ \frac{dy_2}{dt}=\frac{\frac{d\gamma(\bullet)}{dt}\eta(\bullet)-\frac{d\eta(\bullet)}{dt}\left[\left(y_2+\sqrt{\gamma(\bullet)+c_2\left[y_1+\frac{c_1}{\eta(\bullet)}\right]}\right)^2-\gamma(\bullet)\right]}{2\eta(\bullet)\left(y_2+\sqrt{\gamma(\bullet)+c_2\left[y_1+\frac{c_1}{\eta(\bullet)}\right]}\right)}-\\ -\frac{1}{2}\frac{\frac{d\gamma(\bullet)}{dt}-\frac{c_2}{\eta(\bullet)}\frac{d\eta(\bullet)}{dt}\left[y_1+\frac{c_1}{\eta(\bullet)}\right]}{\sqrt{\gamma(\bullet)+c_2\left[y_1+\frac{c_1}{\eta(\bullet)}\right]}}=f_2^2\left(t,y_1,y_2;c_1,c_2,\xi\right)\end{array}\right\}. \tag{93}$$

We obtain the following canonical representation of the initial system $I_x$ described by (90)

$$I_y=\left\{\left\{\begin{array}{l}\Gamma_{x_1}^2(\bullet)=\\ =\{y_1=0\},\\ \Gamma_{x_2}^2(\bullet)=\\ =\{y_2=0\}\end{array}\right\};\left\{\begin{array}{l}\frac{dy_1}{dt}=-\frac{1}{\eta(\bullet)}\frac{d\eta(\bullet)}{dt}y_1=f_1^2\left(t,y_1,y_2;c_1,c_2,\xi\right),\\ \frac{dy_2}{dt}=\frac{\frac{d\gamma(\bullet)}{dt}\eta(\bullet)-\frac{d\eta(\bullet)}{dt}\left[\left(y_2+\sqrt{\gamma(\bullet)+c_2\left[y_1+\frac{c_1}{\eta(\bullet)}\right]}\right)^2-\gamma(\bullet)\right]}{2\eta(\bullet)\left(y_2+\sqrt{\gamma(\bullet)+c_2\left[y_1+\frac{c_1}{\eta(\bullet)}\right]}\right)}-\\ -\frac{1}{2}\frac{\frac{d\gamma(\bullet)}{dt}-\frac{c_2}{\eta(\bullet)}\frac{d\eta(\bullet)}{dt}\left[y_1+\frac{c_1}{\eta(\bullet)}\right]}{\sqrt{\gamma(\bullet)+c_2\left[y_1+\frac{c_1}{\eta(\bullet)}\right]}}=f_2^2\left(t,y_1,y_2;c_1,c_2,\xi\right)\end{array}\right\}\right\}, \tag{94}$$

where $\Gamma_{x_i}^2(\bullet)=\Gamma_{x_i}^2\left(t,y_1,y_2;c_i,\xi\right),i\in(1,2)$.

Ultimately, we have

$$I_x\xleftrightarrow{\overline{\varphi}\left(\overline{\varphi}^{-1}\right)}I_y\ . \tag{95}$$

To use analytical criteria of the classificational stability of the typical fibers of the fiber bundles $\Gamma^2_{x_1}(\bullet)$ and $\Gamma^2_{x_2}(\bullet)$, we need to derive $\frac{\partial f_1^2(t,y_1,y_2;c_1,c_2,\xi)}{\partial y_1}$ and $\frac{\partial f_2^2(t,y_1,y_2;c_1,c_2,\xi)}{\partial y_2}$. We have

$$\begin{aligned}\frac{\partial f_1^2(t,y_1,y_2;c_1,c_2,\xi)}{\partial y_1}&=-\frac{1}{\eta(\bullet)}\frac{d\eta(\bullet)}{dt},\\ \frac{\partial f_2^2(t,y_1,y_2;c_1,c_2,\xi)}{\partial y_2}&=-\frac{1}{2\eta(\bullet)W(\bullet)^2}\left\{\frac{d\eta(\bullet)}{dt}W(\bullet)^2+\frac{d\gamma(\bullet)}{dt}\eta(\bullet)+\frac{d\eta(\bullet)}{dt}\gamma(\bullet)\right\},\end{aligned} \tag{96}$$

where $W(\bullet)=y_2+\sqrt{\gamma(\bullet)+c_2\left[y_1+\frac{c_1}{\eta(\bullet)}\right]}$ .

On the 2-dimensional plane $\{y_1=0\}\in R^3_{t,y_1,y_2}$ and along the straight line being the axis $t$ $\{y_1=0\}\cap\{y_2=0\}\in R^3_{t,y_1,y_2}$ the partial derivative $\frac{\partial f_1^2(t,y_1,y_2;c_1,c_2,\xi)}{\partial y_1}$ takes the same form, namely

$$\left.\frac{\partial f_1^2(t,y_1,y_2;c_1,c_2,\xi)}{\partial y_1}\right|_{y_1=0}=\left.\frac{\partial f_1^2(t,y_1,y_2;c_1,c_2,\xi)}{\partial y_1}\right|_{y_1=0,y_2=0}=-\frac{1}{\eta(\bullet)}\frac{d\eta(\bullet)}{dt}. \tag{97}$$

But on the 2-dimensional plane $\{y_2=0\}\in R^3_{t,y_1,y_2}$ and along $\{y_1=0\}\cap\{y_2=0\}\in R^3_{t,y_1,y_2}$ the partial derivative $\frac{\partial f_2^2(t,y_1,y_2;c_1,c_2,\xi)}{\partial y_2}$ takes the different forms

$$\left.\frac{\partial f_2^2(t,y_1,y_2;c_1,c_2,\xi)}{\partial y_2}\right|_{y_2=0}=-\frac{1}{2}\left\{\frac{1}{\eta(\bullet)}\frac{d\eta(\bullet)}{dt}+\frac{\frac{d\gamma(\bullet)}{dt}+\frac{1}{\eta(\bullet)}\frac{d\eta(\bullet)}{dt}\gamma(\bullet)}{\gamma(\bullet)+c_2\left[y_1+\frac{c_1}{\eta(\bullet)}\right]}\right\}, \tag{98}$$

$$\left.\frac{\partial f_2^2(t,y_1,y_2;c_1,c_2,\xi)}{\partial y_2}\right|_{y_1=y_2=0}=-\frac{1}{2}\left\{\frac{1}{\eta(\bullet)}\frac{d\eta(\bullet)}{dt}+\frac{\frac{d\gamma(\bullet)}{dt}+\frac{1}{\eta(\bullet)}\frac{d\eta(\bullet)}{dt}\gamma(\bullet)}{\gamma(\bullet)+\frac{c_1c_2}{\eta(\bullet)}}\right\}. \tag{99}$$

Suppose

$$\eta\left(t,\xi_1\right)=e^{\xi_1 t},$$
$$\gamma\left(t,\xi_2,\xi_3,\xi_4\right)=\xi_2+\xi_3\sin\left(\xi_4 t\right) \tag{100}$$

Let us define the domain for $\left(c_1,c_2\right)$ with the following inequalities

$$\Omega=\begin{Bmatrix}-1\le c_1\le 1\\ 0\le c_2\le 1\end{Bmatrix}. \tag{101}$$

Then we get

$$\left.\frac{\partial f_1^2\left(t,y_1,y_2;c_1,c_2,\xi\right)}{\partial y_1}\right|_{y_1=0,y_2=0}=-1, \tag{102}$$

$$\left.\frac{\partial f_2^2\left(t,y_1,y_2;c_1,c_2,\xi\right)}{\partial y_2}\right|_{y_2=0}=-\frac{1}{2}\cdot\left(1+\frac{2+\sin\left(t\right)+\cos\left(t\right)}{2+\sin\left(t\right)+c_2\left[y_1+c_1e^{-t}\right]}\right), \tag{103}$$

$$\left.\frac{\partial f_2^2\left(t,y_1,y_2;c_1,c_2,\xi\right)}{\partial y_2}\right|_{y_1=y_2=0}=-\frac{1}{2}\cdot\left(1+\frac{2+\sin\left(t\right)+\cos\left(t\right)}{2+\sin\left(t\right)+c_1c_2e^{-t}}\right) \tag{104}$$

If

$$c=\begin{vmatrix}\hat{c}_1\\ \hat{c}_2\end{vmatrix}\in\left\{\begin{vmatrix}0\\ 0\end{vmatrix},\ \begin{vmatrix}0.5\\ 0.5\end{vmatrix},\ \begin{vmatrix}-0.5\\ 0.5\end{vmatrix}\right\} \tag{105}$$

and

$$\xi=\hat{\xi}=\begin{vmatrix}\hat{\xi}_1\\ \hat{\xi}_2\\ \hat{\xi}_3\\ \hat{\xi}_4\end{vmatrix}=\begin{vmatrix}1\\ 2\\ 1\\ 1\end{vmatrix}, \tag{106}$$

then MATLAB simulation of the initial system $I_x$ defined by (90) and its canonical representation $I_y$ defined by (94) with the expressions (103) - (104) gives us the results presented by following graphs.

**Example:** Integral curves at $\xi = (1, 2, 1, 1)$: $x_1 = -c_1 \exp(-(t) + \mathrm{sqrt}(c_1 c_2 \exp(-t) + 2 + \sin(t))$,
$x_2 = \mathrm{sqrt}(c_1 c_2 \exp(-(t) + 2 + \sin(t))$

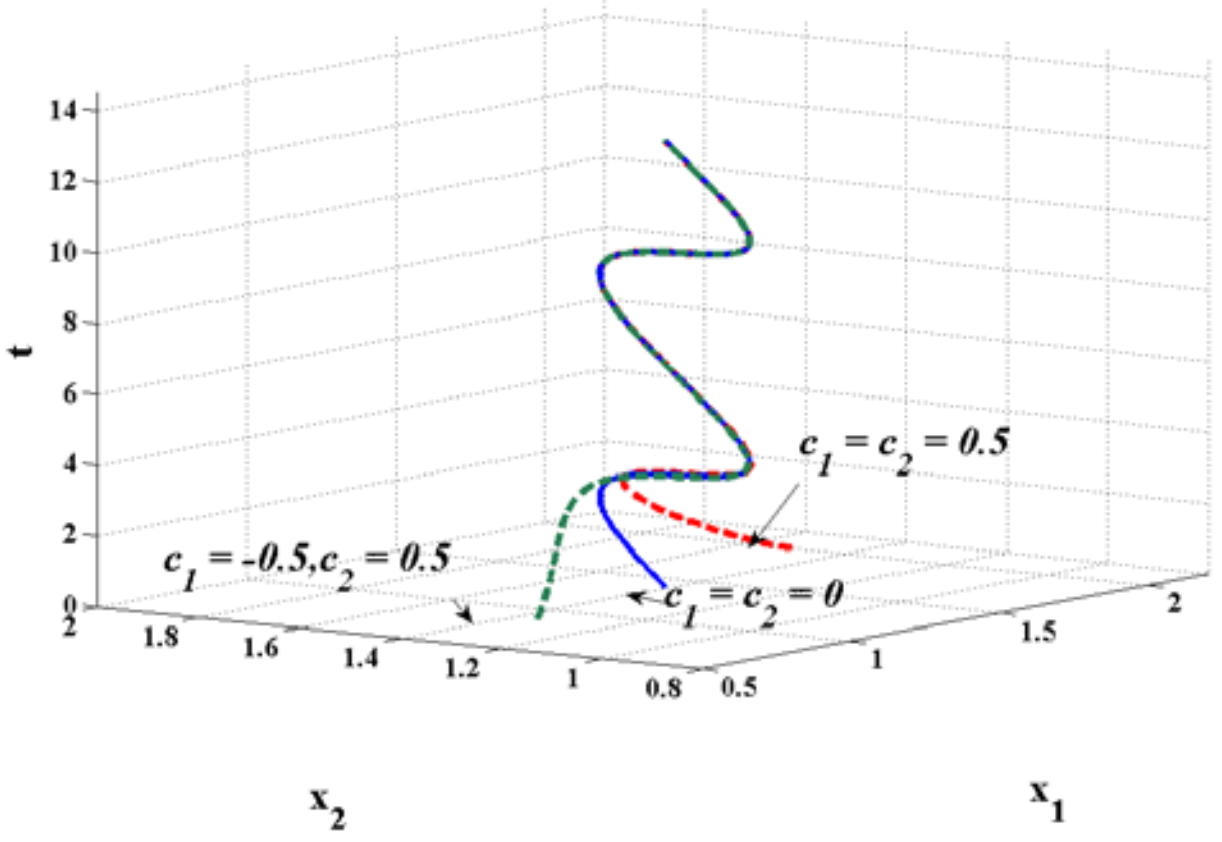

**GRAPH 1**

**Example:** Two specific sections of the corresponding two foliations defined by the first integrals

$g_1(t, x_1, x_2) = (x_1 - x_2)e^t = c_1$ and $g_2(t, x_1, x_2) = (x_2^2 - 2 - \sin(t))/(x_1 - x_2) = c_2$ by the hyperplanes $(c_1 = 0.5)$, $(c_2 = 0.5)$

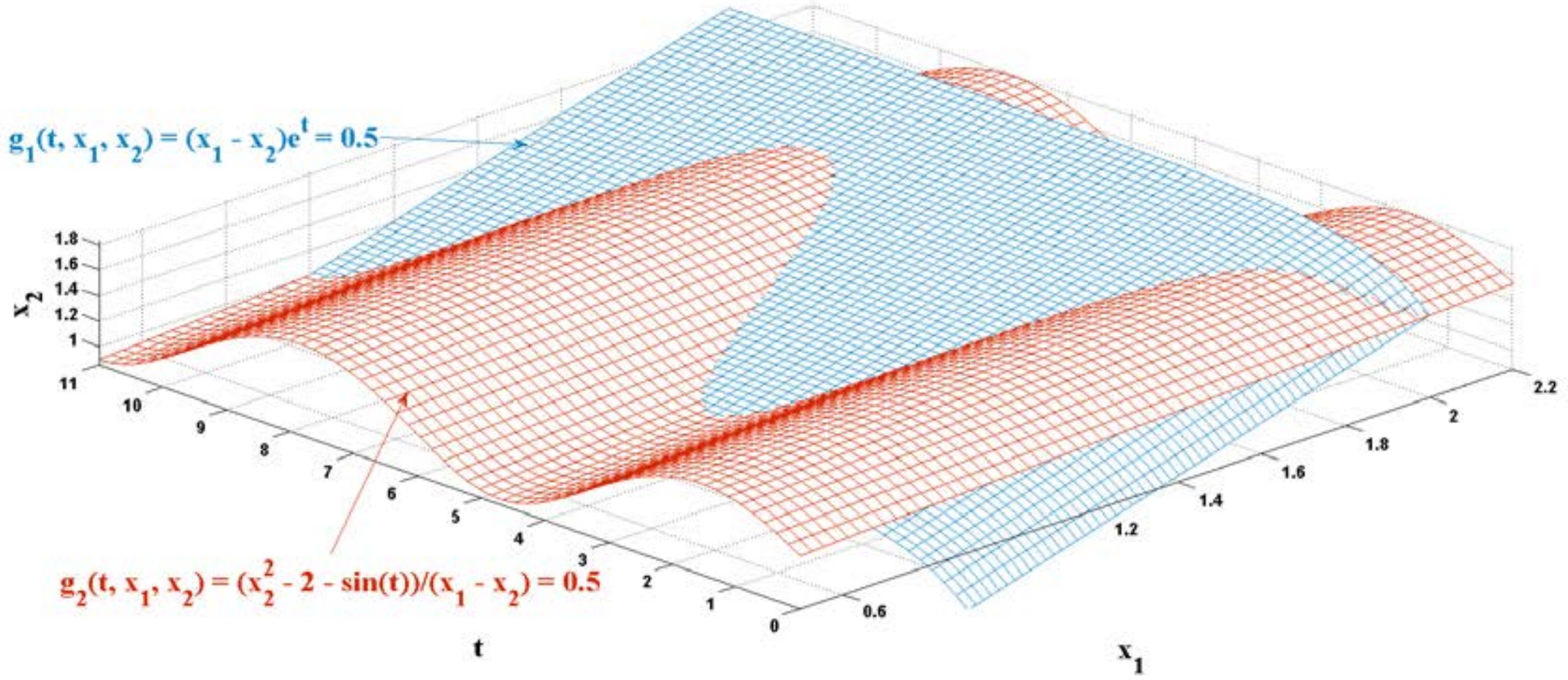

**GRAPH 2**

**EXAMPLE:** The 3D-graph of the expression of $\partial \mathbf{f}_2^2(\mathbf{t}, \mathbf{y}_1, \mathbf{y}_2; \mathbf{c}, \xi) / \partial \mathbf{y}_2 \big|_{\mathbf{y}_2 = 0}$ for the analytical criterion at $\xi = (\xi_1, \xi_2, \xi_3, \xi_4) = \mathbf{(1, 2, 1, 1)}$, $\mathbf{c} = (\mathbf{c}_1, \mathbf{c}_2) \in$ **[(0, 0)** *in blue,* ***(-0.5, 0.5)*** *in green,* ***(0.5, 0.5)*** *in red*].

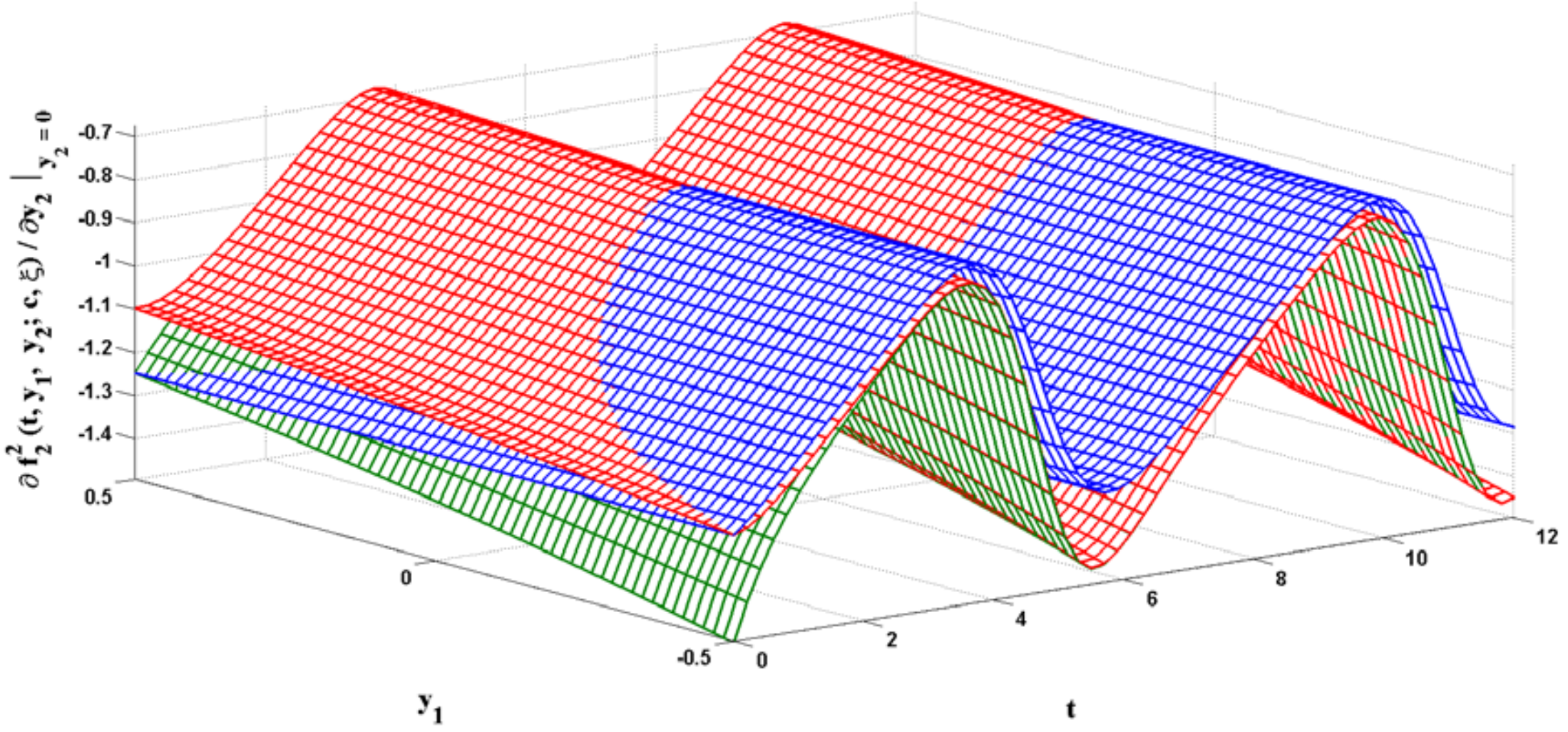

**GRAPH 3**

**EXAMPLE:** The graphs of the expression of $\partial \mathbf{f}_2^2(\mathbf{t}, \mathbf{y}_1, \mathbf{y}_2; \mathbf{c}, \xi) / \partial \mathbf{y}_2 \big|_{\mathbf{y}_2 = \mathbf{y}_1 = 0}$ for the analytical criterion at $\xi = (\xi_1, \xi_2, \xi_3, \xi_4) = \mathbf{(1, 2, 1, 1)}$, $\mathbf{c} = (\mathbf{c}_1, \mathbf{c}_2) \in$ **[(0, 0), (0.5, 0.5), (-0.5, 0.5)].**

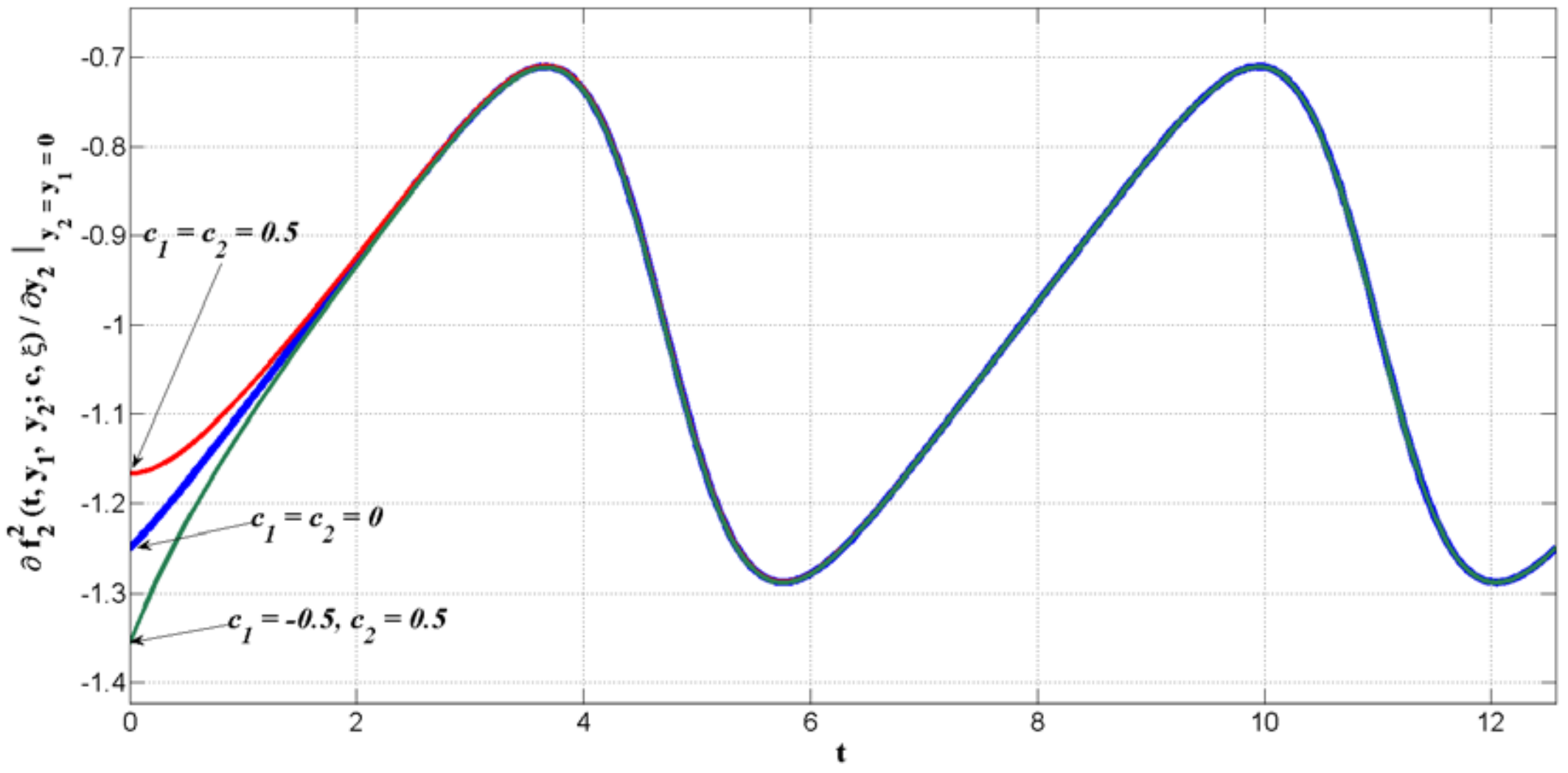

**GRAPH 4**

**EXAMPLE:** The 3D-graph of the surface $\mathbf{S_2 \cap (t = 6) = (s_2 = f_2^2\,(t, y_1, y_2; c, \xi)|_{t=6})}$ being the time section of the 3-dimensional hypersurface $\mathbf{S_2 = (s_2 = f_2^2\,(t, y_1, y_2; c, \xi))}$ for the topological criterion at $\mathbf{\xi = (\xi_1, \xi_2, \xi_3, \xi_4) = (1, 2, 1, 1), c = (c_1, c_2) = (0, 0).}$

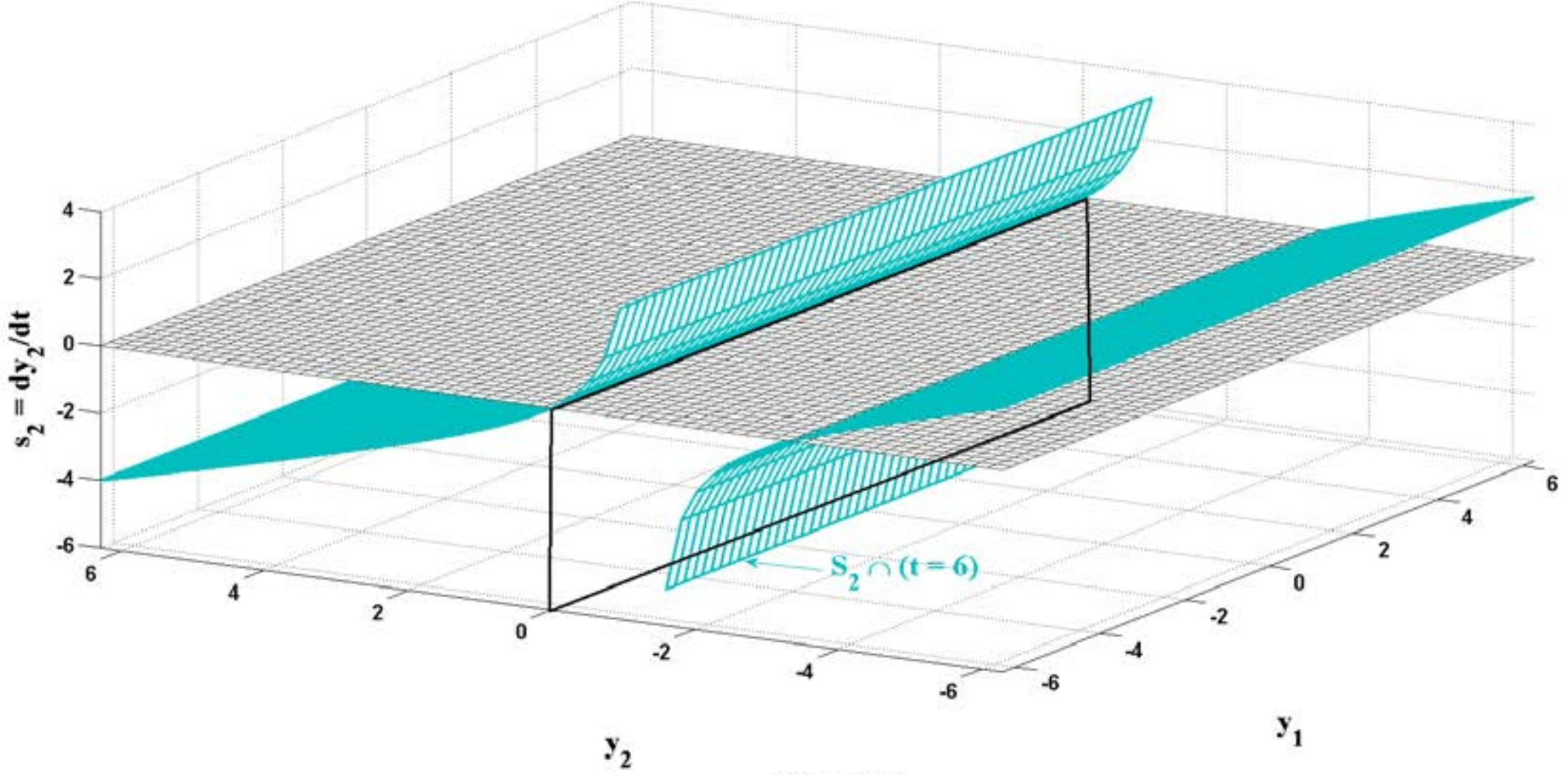

**GRAPH 5**

**EXAMPLE:** The 3D-graphs of the surfaces $\mathbf{S_2 \cap (t = 0)}$ and $\mathbf{S_2 \cap (t = 12)}$ being the time sections of the 3-dimensional hypersurface $\mathbf{S_2 = (s_2 = f_2^2\,(t, y_1, y_2; c, \xi))}$ by the hyperplanes **(t=0)** and **(t=12)** for the topological criterion at $\mathbf{\xi = (\xi_1, \xi_2, \xi_3, \xi_4) = (1, 2, 1, 1), c = (c_1, c_2) = (0.5, 0.5).}$

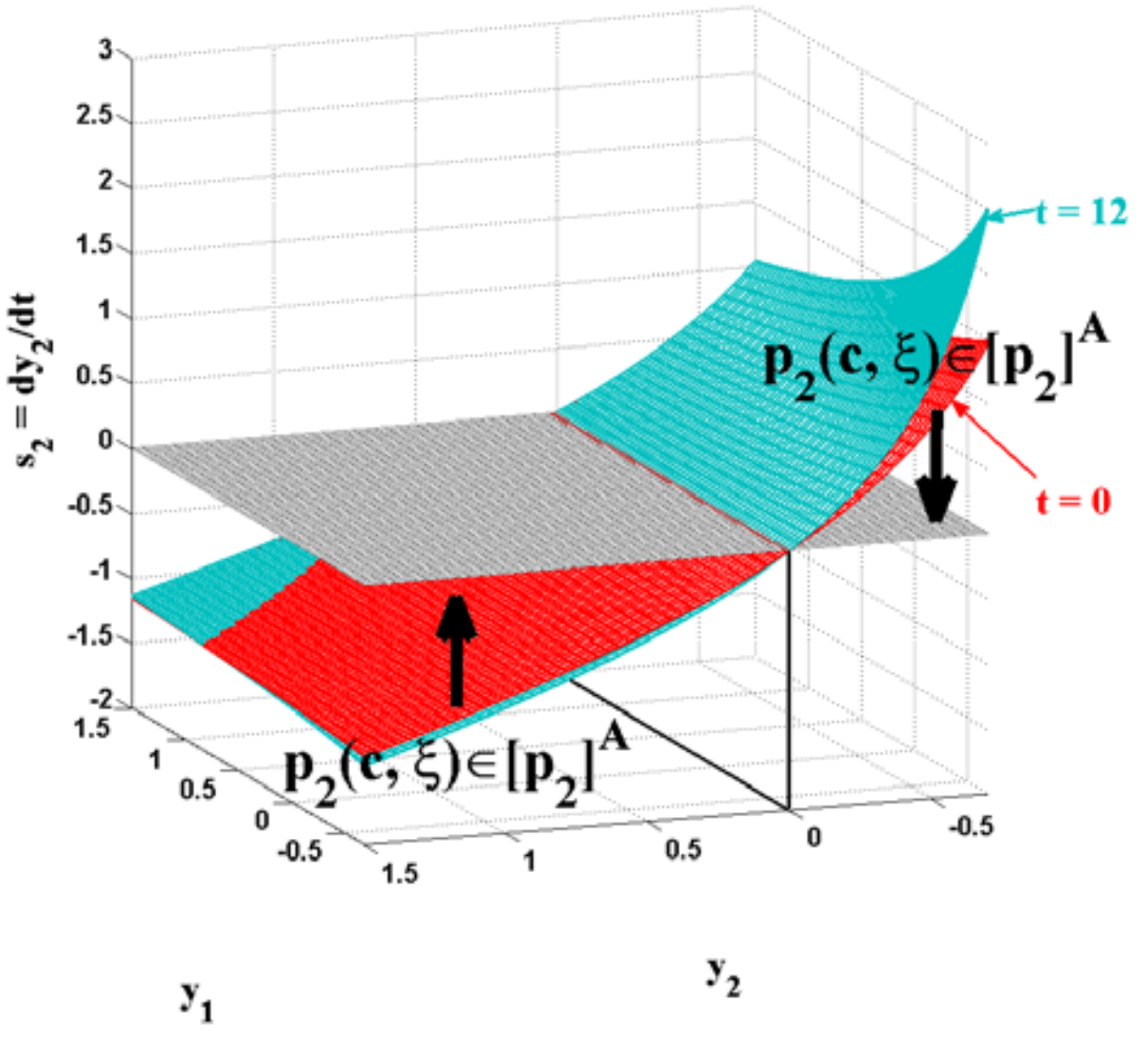

**GRAPH 6**

Since according to the conditions (100), (105) and (106) the canonizing diffeomorphism (92) is uniformly convergent, it is quite evident that under our conditions for the value of $\xi$ and the values of $(c_1, c_2) \in \Omega$, there exists the canonical positive definite auxiliary function $W(y)$ of Lyapunov functions $V(t, y)$ for the canonical system $I_y$ in the form

$$W(y) = y_1^2 + y_2^2 \tag{107}$$

that generates the canonical positive definite Lyapunov function

$$V(t, y) = W(y)\left(1 + e^{-t}\right) = \left(y_1^2 + y_2^2\right)\left(1 + e^{-t}\right) \ , \tag{108}$$

ensuring the asymptotic stability of all its integral curves defined by the above-mentioned assumptions. According to Theorem 1, this function also guaranties the asymptotical stability of the corresponding integral curves or typical 1-dimensional fibers of the initial system $I_x$.

***Remark 5.*** *It is evident that*

$$V(t, y)\Big|_{y_1 = y_2 = 0} = \left(y_1^2 + y_2^2\right)\left(1 + e^{-t}\right) = 0, V(t, y)\Big|_{\forall\left(y_1, y_2\right) \neq \left(0,0\right)} > 0. \tag{109}$$

*The total derivative of* $V(t, y)$ *is*

$$\frac{dV(t,y)}{dt} = -e^{-t}\left(y_1^2 + y_2^2\right) - 2y_1^2 + 2y_2\ f_2^2\left(t, y_1, y_2; c_1, c_2, \hat{\xi}\right)\Big|_{c \in \Omega, \hat{\xi} = \left(1,2,1,1\right)}, \tag{110}$$

*where* $\left.\dfrac{dV(t, y)}{dt}\right|_{y_1 = y_2 = 0} = 0$.

*Now we will show that the total derivative is negative definite. Two first terms are beyond any doubts. As to the term* $2y_2\ f_2^2\left(t, y_1, y_2; c_1, c_2, \hat{\xi}\right)\Big|_{c \in \Omega, \hat{\xi} = \left(1,2,1,1\right)}$, *it is also negative for all* $(y_1, y_2) \neq (0,0)$ *because the covering mapping* $p_2\left(c, \hat{\xi}\right)$ *belongs to the class of* $\mathbf{A}$ *– mappings, that is* $p_2\left(c, \hat{\xi}\right) \in [p_2]^{\mathbf{A}}$. *The last statement means that for* $\xi = \hat{\xi} = (1,2,1,1)$ *and* $\forall c \in \Omega$

$$f_2^2\left(t, y_1, y_2; c_1, c_2, \hat{\xi}\right)\Big|_{y_2 = 0} \equiv 0 \tag{111}$$

*and*

$$f_2^2\left(t,y_1,y_2;c_1,c_2,\hat{\xi}\right)\Big|_{y_2>0}<0,\ f_2^2\left(t,y_1,y_2;c_1,c_2,\hat{\xi}\right)\Big|_{y_2<0}>0 \qquad (112)$$

*in the quite large domain* $\{1.5>y_1>-0.5;1.5>y_2>-0.5;\infty>t\geq 0\}$ *(see Graph 6 as an illustration). The values of the vector of parameters* $\xi$ *can be varied rather extensively from the chosen one of* $\hat{\xi}=(1,2,1,1)$ *but all the properties of asymptotical stability will be preserved and the qualitative results of the conducted analysis will remain unchangeable.*

## CONCLUSION

We have achieved the goals

**1)** to make the geometric-topological representation of nonlinear non-autonomous parametric differential inclusions in the framework of fiber bundles and foliations;

**2)** to develop the general procedure of the utilization of Lyapunov functions for the parametric differential inclusions through the canonizing diffeomorphisms defined by the full sets of their parametric first integrals;

**3)** to investigate

- **a)** the asymptotic stability in the sense of Lyapunov of particular solutions of free dynamic systems as an example of local stability and
- **b)** the global asymptotic stability of the parametric differential inclusions and free dynamic systems as their restrictions.

Let us pay more detailed attention to the points **1)** and **2)**.

First, the geometric-topological structure of the parametric differential inclusions is hierarchical. Each point $\hat{\xi}$ of the manifold $\Xi$ of first-grade or absolutely independent vector of parameters $\xi$ generates a free dynamic system within some given inclusion. The extended phase space or the motion space of the system contains $n$ 1-codimensional smooth foliations, the leaves of which intersect each other transversely. These leaves are the elementary “bricks” of the space. The intersection of $n$ leaves of different foliations gives us an integral curve. The system has one more absolutely independent vector of parameters, namely $c=(c_1,...,c_n)\in R^1\underbrace{\times...\times}_{(n-1)\ times}R^1=R^n$ that is the right-hand sides of the vector equation (12) for the full set of first integrals. Generally it belongs to some $n$-dimensional manifold $C\subseteq R^n$. It has also the second-grade independent vector of parameter $x_0$ being the initial points of integral curves and depending on $c$. The vector of parameters $c$ creates the quotient space $X_{t_0} / L_{x_i}\left(\hat{c}_i, \hat{\xi}\right)$ from the manifold $X_{t_0}$ of the initial points of the phase vector $x$, where $L_{x_i}\left(\hat{c}_i, \hat{\xi}\right)$ are $(n-1)$-dimensional manifolds, formed by the intersection of the leaves and the hyperplane $\{t = t_0\}$. The 1-codimensional foliations "stem from" the elements $L_{x_i}\left(\hat{c}_i, \hat{\xi}\right)$ of the quotient space $X_{t_0} / L_{x_i}\left(\hat{c}_i, \hat{\xi}\right)$ of the manifold $X_{t_0}$. The structure of fiber bundles can be easily constructed from the structure of foliations if we consider the manifolds of the vectors of parameters $\Xi$ and $R^1$ or $X_{t_0} / L_{x_i}\left(\hat{c}_i, \hat{\xi}\right)$ as base spaces.

Second, after we have found the canonical form of the representation of the inclusions it has not been difficult to see the obvious fact formulated as the Theorem 1. Unfortunately it confirms our worst apprehensions: the Lyapunov functions are some kind of artificial creatures having no bearings on the intrinsic properties of the concrete parametric differential inclusions expressed through their parametric first integrals and right-hand sides. Now it is clear why constructing or finding Lyapunov functions for some concrete nonlinear non-autonomous parametric differential inclusion of general form is incredibly difficult problem: *neither right-hand sides nor even first integrals give us any hint at how to address the problem*. Moreover, some specific Lyapunov function can be successfully applied with the same successful results to the absolutely different differential inclusions. And on the contrary, a given differential inclusion can have at least an infinite countable set of Lyapunov functions. They can argue that this ascertainment is valid only for the canonical forms of the parametric differential inclusions and if we try to go back to the variables *x* using the inverse canonizing diffeomorphism, which is generated by the vector equation (12) for the parametric first integrals of the initial inclusion (1), then it is possible to receive the Lyapunov functions that will take account of the ones. Even if we are of success on this way, although there are serious doubts about that, it is very difficult to ignore the very strong fact presented by the Theorem 1: *we can do completely without Lyapunov functions, which have proved to be the superfluous tool in the canonical case, to determine the asymptotical stability not even of a particular integral curve of some particular system of differential equations but all the integral curves of the whole parametric differential inclusion (1), within of which the above-mentioned system corresponds just to a single point of m-dimensional Euclidian space*.

The last point that we would like to discuss is the prospects of the Lyapunov second method for the investigation of stability in the light of our discovery. Does it downplay the significance of the method? The answer is: by no means. The downsides of the method have been known for a long time. What was not known this was the reason why? Now we are aware of it and this fact entails the understanding that the Lyapunov second method provides us only with sufficient conditions of asymptotic stability. This means there is a good chance to encounter the following unpleasant phenomenon: directly applying the main theorem of the Lyapunov second method to a given parametric differential inclusion in its canonical form, which is actually asymptotically stable, does not evince any satisfactory results on the existence of this property in the inclusion. The typical case of such a situation is the presence of mosaic attractors, introduced in [6], in the motion spaces of dynamic systems. The structure of their mosaic patterns can be very complex. The Lyapunov functions are absolutely insensitive to asymptotic stability created by them because they are not able to detect the presence of mosaic attractors in motion spaces. Thus, parametric differential inclusions and their restrictions as dynamic systems or integral curves can be asymptotically stable but it is impossible to apply the Lyapunov second method in such way that the one will confirm this fact. There are two ways out of this situation. The first one is to modify the method in order to add to its "credentials" the feature able to discern mosaic attractors. The second way is to use the Poincare approach developed in [6] to investigate the problems of stability.